 \documentclass[11pt,a4paper]{article}
 \usepackage{amsmath}
\usepackage{cite}
\usepackage{graphicx}
\usepackage{amsfonts}
\usepackage{amssymb}
\input epsf

\def\br{\begin{eqnarray}}
\def\er{\end{eqnarray}}
\def\be{\begin{equation}}
\def\ee{\end{equation}}
\def\({\left(}
\def\){\right)}
\def\rlx{\relax\leavevmode}
\def\IR{\rlx\hbox{\rm I\kern-.18em R}}

\def\a{\alpha}
\def\s{\sigma}

	\newcommand{\ba}{\begin{array}}
	\newcommand{\ea}{\end{array}}
	\newcommand{\beqa}{\begin{equation}\begin{array}{rcl}}
	\newcommand{\eeqa}[1]{\end{array}\label{#1}\end{equation}}

\def\IZ{\rlx\hbox{\sf Z\kern-.4em Z}}
\def\IR{\rlx\hbox{\rm I\kern-.18em R}}
\def\IC{\rlx\hbox{\,$\inbar\kern-.3em{\rm C}$}}
\def\one{\hbox{{1}\kern-.25em\hbox{l}}}

\renewcommand{\theequation}{\thesection.\arabic{equation}}
\def\IZ{\rlx\hbox{\sf Z\kern-.4em Z}}
\def\IR{\rlx\hbox{\rm I\kern-.18em R}}
\def\IC{\rlx\hbox{\,$\inbar\kern-.3em{\rm C}$}}
\def\one{\hbox{{1}\kern-.25em\hbox{l}}}
     
\begin{document}
     
\newcommand{\sect}[1]{\setcounter{equation}{0}\section{#1}}
\renewcommand{\theequation}{\thesection.\arabic{equation}}



     
     
     
     

\newcommand{\m}{ \mu}
\newcommand{\la}{ \lambda}
\newcommand{\C}{ C_0 }
\newcommand{\p}{ \varphi}

\newcommand{\ka}{ \kappa}

\newcommand{\e}{ \varepsilon}

\newcommand{\spa}{$\spadesuit$}
\newcommand{\bul}{$\bullet$}

\newcommand{\ir}{{\mathrm{IR}}}
\newcommand{\uv}{{\mathrm{UV}}}

\newcommand{\ml}{{\mathrm{ml}}}
\newcommand{\ms}{{\mathrm{ms}}}
\newcommand{\ns}{{\mathrm{n.s.}}}

\newcommand{\eff}{{\mathrm{eff}}}

\newcommand{\til}[1]{\tilde{#1}}

\def\PRL#1#2#3{{\sl Phys. Rev. Lett.} {\bf#1} (#2) #3}
\def\NPB#1#2#3{{\sl Nucl. Phys.} {\bf B#1} (#2) #3}
\def\NPBFS#1#2#3#4{{\sl Nucl. Phys.} {\bf B#2} [FS#1] (#3) #4}
\def\CMP#1#2#3{{\sl Commun. Math. Phys.} {\bf #1} (#2) #3}
\def\PRD#1#2#3{{\sl Phys. Rev.} {\bf D#1} (#2) #3}
\def\PRB#1#2#3{{\sl Phys. Rev.} {\bf B#1} (#2) #3}

\def\PLA#1#2#3{{\sl Phys. Lett.} {\bf #1A} (#2) #3}
\def\PLB#1#2#3{{\sl Phys. Lett.} {\bf #1B} (#2) #3}
\def\JMP#1#2#3{{\sl J. Math. Phys.} {\bf #1} (#2) #3}
\def\PTP#1#2#3{{\sl Prog. Theor. Phys.} {\bf #1} (#2) #3}
\def\SPTP#1#2#3{{\sl Suppl. Prog. Theor. Phys.} {\bf #1} (#2) #3}
\def\AoP#1#2#3{{\sl Ann. of Phys.} {\bf #1} (#2) #3}
\def\PNAS#1#2#3{{\sl Proc. Natl. Acad. Sci. USA} {\bf #1} (#2) #3}
\def\RMP#1#2#3{{\sl Rev. Mod. Phys.} {\bf #1} (#2) #3}
\def\PR#1#2#3{{\sl Phys. Reports} {\bf #1} (#2) #3}
\def\AoM#1#2#3{{\sl Ann. of Math.} {\bf #1} (#2) #3}
\def\UMN#1#2#3{{\sl Usp. Mat. Nauk} {\bf #1} (#2) #3}
\def\FAP#1#2#3{{\sl Funkt. Anal. Prilozheniya} {\bf #1} (#2) #3}
\def\FAaIA#1#2#3{{\sl Functional Analysis and Its Application} {\bf #1} (#2)
#3}
\def\BAMS#1#2#3{{\sl Bull. Am. Math. Soc.} {\bf #1} (#2) #3}
\def\TAMS#1#2#3{{\sl Trans. Am. Math. Soc.} {\bf #1} (#2) #3}
\def\InvM#1#2#3{{\sl Invent. Math.} {\bf #1} (#2) #3}
\def\LMP#1#2#3{{\sl Letters in Math. Phys.} {\bf #1} (#2) #3}
\def\IJMPA#1#2#3{{\sl Int. J. Mod. Phys.} {\bf A#1} (#2) #3}
\def\AdM#1#2#3{{\sl Advances in Math.} {\bf #1} (#2) #3}
\def\RMaP#1#2#3{{\sl Reports on Math. Phys.} {\bf #1} (#2) #3}
\def\IJM#1#2#3{{\sl Ill. J. Math.} {\bf #1} (#2) #3}
\def\APP#1#2#3{{\sl Acta Phys. Polon.} {\bf #1} (#2) #3}
\def\TMP#1#2#3{{\sl Theor. Mat. Phys.} {\bf #1} (#2) #3}
\def\JPA#1#2#3{{\sl J. Physics} {\bf A#1} (#2) #3}
\def\JSM#1#2#3{{\sl J. Soviet Math.} {\bf #1} (#2) #3}
\def\MPLA#1#2#3{{\sl Mod. Phys. Lett.} {\bf A#1} (#2) #3}
\def\JETP#1#2#3{{\sl Sov. Phys. JETP} {\bf #1} (#2) #3}
\def\JETPL#1#2#3{{\sl  Sov. Phys. JETP Lett.} {\bf #1} (#2) #3}
\def\PHSA#1#2#3{{\sl Physica} {\bf A#1} (#2) #3}
\def\PHSD#1#2#3{{\sl Physica} {\bf D#1} (#2) #3}

\begin{titlepage}
\vspace*{-2 cm}
\noindent
\begin{flushright}
\end{flushright}

\vskip 1 cm
\begin{center}
{\Large\bf Self-Duality from New Massive Gravity Holography} \vglue 1  true cm

  {U. Camara dS}$^{*}$\footnote {e-mail: ulyssescamara@gmail.com}, {C.P. Constantinidis}$^{*}$\footnote {e-mail: cpconstantinidis@gmail.com}, { A.L. Alves Lima}$^{*}$\footnote {e-mail: andrealves.fis@gmail.com} and { G.M.Sotkov}$^{*}$\footnote {e-mail: sotkov@cce.ufes.br, gsotkov@yahoo.com.br}\\

\vspace{1 cm}

${}^*\;${\footnotesize Departamento de F\'\i sica - CCE\\
Universidade Federal de Espirito Santo\\
29075-900, Vitoria - ES, Brazil}\\

\vspace{5 cm}

\end{center}

\normalsize
\vskip 0.5cm

\begin{center}
{ {\bf ABSTRACT}}\\
\end{center}

\vspace{0.5cm}

The holographic renormalization group (RG) flows in certain self-dual two dimensional QFT's models are studied. They are constructed as holographic  duals to specific New Massive 3d Gravity (NMG) models coupled to scalar matter with ``partially self-dual'' superpotentials. The standard holographic RG constructions allow us to derive the exact form of their $\beta$- functions in terms of the corresponding NMG's domain walls solutions. By imposing  invariance  of the  free energy, the central function and of the anomalous dimensions under specific matter field's duality transformation, we have found the conditions  on the superpotentials of two different NMG's models, such that their dual 2d QFT's are related by a simple strong-weak coupling transformation.

\vspace{0.5 cm} 

\end{titlepage}

\tableofcontents 
\setcounter{equation}{0}
\section{Introduction}

In the lack of small parameters, the concepts and methods of the strong-weak coupling duality \cite{seiberg,seibit} are known to be the main tool for the description of relevant physical phenomena, and for the derivation of non-perturbative strong coupling results.
In all the known examples of self-dual (supersymmetric) QFT$_d$'s (with $d=2,3,4$), this  duality is realized as an inversion transformation or, more generally, as fractional linear transformations of the couplings belonging to certain discrete subgroups of SL(2,C), 
which leave invariant the corresponding partition functions \cite{seiberg,seibit,haro,cardy-itz}. The gauge/gravity  duality \cite{malda,witt,maldadu,ooug} on the other hand, together with  the holographic  Renormalization Group (RG) \cite{VVB,rg}, establish an equivalence relation between certain limits of the (semi-)classical $d$-dimensional gravity models and the strong-coupling regime of  $(d-1)$-dimensional gauge theories. According to the off-critical holographic RG version of the AdS/CFT correspondence, the  QFT$_{d-1}$'s  dual to certain asymptotically AdS$_d$ geometries of domain wall (DW's) type \cite{rg}  may involve --- together with the original gauge strong coupling --- a few other  relevant or/and marginal couplings. These models can be also realized as certain conformal perturbations around a given CFT, thus defining  non-conformal theories called pCFT's \cite{x, cardy}. We are interested in the specific holographic features of such dual, non-conformal QFT's, in the case when they  belong to the family of the self-dual theories w.r.t. one (or a few) of these couplings. More precisely,  we shall address the question of how can one derive the ``holographic gravitational'' counterparts of  certain duality symmetries, such as the above mentioned inversions and  fractional linear  transformations. We consider the 3-dimensional New Massive Gravity (NMG) model \cite{1}, coupled to scalar self-interacting matter, and will look for the specific restrictions to be imposed on the form of the matter superpotential in order to ensure  the strong-weak coupling self-duality of the corresponding two-dimensional pCFT$_2$, constructed by the methods of the NMG holography \cite{sinha,nmg,oldholo,iran}.

The recent progress in the understanding of the t'Hooft  limits ($N, \, k \rightarrow \infty$ but finite $\frac{N}{N+k}$) of certain cosets of SU$(N)_k$ WZW models (as for example the $W_N$ minimal models) as an appropriate higher spin extension of the 3d Einstein gravity \cite{gaber, gab, minmod} has  renewed the interest in the identification of appropriate limits of the most famous family of CFT$_2$'s --- the BPZ and  the  Liouville  minimal models \cite{bpz} --- as holographic duals of certain extended 3d gravity models \cite{sinha, nmg, oldholo, iran}. There exists an indication that the holographic description of these CFT$_2$'s in the case of relatively large, but \emph{finite central charges}, can be achieved  by considering the  quantum 3d gravity contributions beyond the (semi-)classical one \cite{gab}, or/and of certain ``higher curvature" extensions of the Einstein gravity, including  powers of the curvature and of the Ricci tensor at the classical level. The simplest model of such an extended 3d gravity is given by the following  action, called  New Massive Gravity
 \footnote{One may consider the new $R^2$-type terms  as one loop counter-terms appearing in the perturbative quantization of 3d Einstein gravity with $\Lambda<0$.}
\cite{1}:             
\begin{eqnarray}
 && S_{{\mathrm{NMG}}}(g_{\mu\nu};\kappa,\Lambda)= \frac{1}{\kappa^2}\int d^3x\sqrt{-g}\left[\epsilon R+ \frac{1}{m^2} \Big(R_{\mu\nu}R^{\mu\nu}-\frac{3}{8}R^2\Big)-2\Lambda \right],\label{acao} \\
&&  \kappa^2=16\pi G,\;\; \epsilon=\pm1. \nonumber
\end{eqnarray}
At the linearised level, it  describes  a massive graviton with  two polarizations. As it was shown by Bergshoeff, Hohm and Townsend (BHT) \cite{1}, the above model turns out to be \emph{unitary} consistent (ghost free)  for both choices, $\epsilon=\pm1$, of the ``right'' and ``wrong'' signs of the $R$-term, under certain restrictions on the values of the cosmological constant $\Lambda=-m^2\lambda$, as for example\cite{more}:
\be 
  -1\le\lambda<0 \; , \quad\quad  \epsilon = -1 \, ,\quad \;\;   m^2<0 \; . \label{bhtun} 
\ee
in the case of the negative $\lambda$ \emph{BHT-unitary window}.

An important feature of the central charges of the  CFT$_2$'s  dual to these NMG models is the presence of a particular $m^2$-dependent term \cite{more,china}:
\begin{eqnarray}
 c_{nmg} =\frac{3\epsilon L}{2l_{pl}}\left(1+\frac{L_{gr}^2}{L^2}\right) \; , \quad L_{gr}^2=\frac{1}{2\epsilon m^2}\gg l_{pl}^2 \; , \quad \Lambda_{\eff}=-\frac{1}{L^2}=-2m^2(\epsilon\pm\sqrt{1+\lambda}). \label{ch}
\end{eqnarray} 
Compared to the  standard 3d Einstein gravity case, which one resumes to in the  $m^2\rightarrow\infty$ limit, the above central charges yield a remarkable new \emph{self-duality} property: $c_{nmg} (L)=c_{nmg} \left( L_{gr}^2 / L \right)$, coinciding with the well  known  ``$b$ to $1/b$" duality of the (exact, non-perturbative) central charges  $c^{\pm}(b)=1 \pm 6(b\pm\frac{1}{b})^2$ of the  Liouville ($c^+$) and of the BPZ ($c^-$) minimal models\cite{bpz,azz}. It is then natural to expect that appropriate perturbations of these CFT's give rise to certain strong-weak coupling  self-dual non-conformal $pCFT_2$'s we are interested in. Although the proper identification of the CFT$_2$'s dual to NMG model (\ref{acao}), is not yet fully understood, the \emph{off-critical} AdS/CFT methods based on the DW's solutions of NMG model coupled to massive self-interacting scalar field with an action \cite{nmg,pos}:
\begin{eqnarray}
&& S_{{\mathrm{NMGm}}}(g_{\mu\nu},\sigma;\kappa,\Lambda) =  \frac{1}{\kappa^2} \int d^3x\sqrt{-g} \left\{ \epsilon R+ \frac{1}{m^2} {\cal K}-\kappa^2 \left(\frac{1}{2} |\vec{\nabla}\sigma|^2+V(\sigma)\right)\right\} ; \label{acaoo}\\
&& {\cal{K}} =R_{\mu\nu}R^{\mu\nu}-\frac{3}{8}R^2, \quad \Lambda=-\frac{\kappa^2}{2}V(\sigma^*),\quad V'(\sigma*)= 0 ; \nonumber
\end{eqnarray}
 as well as the  NMG holographic RG results related to them \cite{nmg,oldholo},  provide  the necessary tools for the selection of the conditions on the NMG-matter interactions which lead to such \emph{self-dual} pCFT$_2$'s. 

Our main  result  consists in the explicit construction of the duality transformations between \emph{pairs of 3d NMG-matter models} (\ref{acaoo}), whose holographic 2d images represent specific strong-weak coupling transformations which keep invariant the  free energy, the corresponding $C$-function and the anomalous dimensions of their pCFT$_2$ duals. We also derive the explicit form of a partially self-dual matter superpotential (with all the vacua within the negative BHT-unitary window (\ref{bhtun})) giving rise to a holographic, self-dual, pCFT$_2$ model, presenting both strong- and weak-coupling phases and critical points. The practical importance of the concept of \emph{partial self-duality}, introduced in Sect.\ref{Examples of dual and self-dual}., is that it  provides an efficient method for the identification of such holographic pCFT$_2$'s of a given exact $\beta$-function, by comparing the results concerning its weak-coupling phases with the standard and well known perturbative CFT$_2$'s calculations around a given (weak-coupling) critical point \cite{cardy, x, fat, gms}.


\setcounter{equation}{0}
\section{NMG holography}

The models involved in the ``boundary'' QFT$_2$'s part of the \emph{off-critical} AdS$_3$/CFT$_2$ correspondence \cite{gub}  are usually identified as certain CFT$_{2}$'s, perturbed by marginal or/and relevant operators that break the conformal symmetry of it's Poincar\'e subgroup:
\begin{eqnarray}
S_{pCFT_2}^{ren}(\sigma)=S_{CFT_2}^{\uv}+\sigma(L_*)\int \! d^2x \; \Phi_{\sigma}(x^i) .   \label{eq28}
\end{eqnarray}
The scale-radial duality \cite{VVB,rg} allows to further identify the ``running'' coupling constant $\sigma(L_*)$ of the pCFT$_2$   with the scalar field  $\sigma(z)$, and the RG scale $L_*$ with the scale factor $e^{\varphi(z)}$ of the DW's solutions of the bulk gravity coupled to scalar matter, as follows:
\begin{eqnarray}
ds^2=dz^2+e^{\varphi(z)}(dx^2-dt^2),\quad\quad
\sigma(x^i,z)\equiv\sigma(z),\quad \ L_*=l_{pl}e^{-\varphi/2} .      \label{intro1} 
\end{eqnarray}
The main ingredients of the NMG holography -- the NMG's vacua  and DW's solutions, the values of the central charges  of the conjectured dual $CFT_2$'s and the holographic RG flows -- were extensively studied  by different methods \cite{1,2,3,nmg,oldholo,8}. As is well known from
the  example of Einstein gravity \cite{6} ,the properties and the proper existence of the holographic RG flows  in its 2d dual QFT$_2$, strongly depend on the form of bulk matter interactions. If they permit DW's solutions relating two unitary NMG vacua of different $\lambda_A$, then we might have  massless RG flows in the dual pCFT$_2$. The explicit construction of all the DWs solutions of the corresponding second order system of equations:
\begin{eqnarray}
&&\ddot{\sigma}+\dot{\sigma}\dot{\varphi}-V'(\sigma)=0\nonumber\\
&&\ddot{\varphi}\Big(1-\frac{\dot{\varphi}^2}{8\epsilon m^2}\Big)+\frac{1}{2}\dot{\varphi}^2\Big(1-\frac{\dot{\varphi}^2}{16\epsilon m^2}\Big)+\epsilon\kappa^2\Big(\frac{1}{2}\dot{\sigma}^2+V(\sigma)\Big)=0\nonumber\\
&&\dot{\varphi}^2\Big(1-\frac{\dot{\varphi}^2}{16\epsilon m^2}\Big)+\epsilon\kappa^2(-\dot{\sigma}^2+2V(\sigma))=0\label{eq4}
\end{eqnarray}
is  a rather difficult problem, and in general it requires the use of numerical methods. However, one particular class of such solutions which are ``stable'' and exact can be obtained by introducing an auxiliary function $W(\sigma)$, called superpotential, which allows to reduce the corresponding DW's gravity-matter equations to an specific BPS-like $I^{st}$ order system \cite{3,nmg}:
\begin{eqnarray}
\kappa^2V(\sigma)&=&2(W')^2\Big(1-\frac{\kappa^2W^2}{2\epsilon m^2}\Big)^2-2\epsilon\kappa^2 W^2\Big(1-\frac{\kappa^2 W^2}{4\epsilon m^2}\Big),\nonumber\\
\dot{\varphi}&=&-2\epsilon\kappa W, \quad \dot{\sigma}=\frac{2}{\kappa}W'\Big(1-\frac{\kappa^2W^2}{2\epsilon m^2}\Big) \, , \label{sis}
\end{eqnarray}
where $W'(\sigma)=dW/d\sigma$, $\dot{\sigma}= d\sigma / dz$ etc.
This provides the explicit form of  qualitatively new DW's relating ``old" and ``new" purely NMG vacua, as well as  of the corresponding pCFT$_2$ model's $\beta$-function \cite{nmg}.

Given the form of the superpotential $W(\sigma)$ and the $I^{st}$ order system (\ref{sis}) --- which describes the radial evolution of the NMG's scale factor and of the scalar field $\sigma(z)$ ---, the scale-radial identifications (\ref{intro1}) provide the explicit form of the $\beta$-function of the conjectured dual QFT$_2$ \cite{VVB,rg}:
\begin{eqnarray}
\frac{d\sigma}{dl}=-\beta(\sigma)=\frac{2\epsilon}{\kappa^2}\frac{W'(\sigma)}{W(\sigma)}\bigg(1-\frac{W^2(\sigma)\kappa^2}{2\epsilon m^2}\bigg),\quad\quad l=\ln L_*  \; .   \label{rg}
\end{eqnarray}
The admissible constant solutions $\sigma^*_{A}$ of the above RG equation (\ref{rg}) are defined by the zeros of the $\beta$-function, and they indeed coincide with the NMG-matter models vacua solutions of AdS$_3$ type. The variety of different \emph{non-constant} solutions $\sigma_{ij}=\sigma(l;\sigma^*_{A_i},\sigma^*_{A_j})$
representing the way the coupling constant $\sigma(l)$ of the dual QFT$_2$ is running (with the RG scale $L_*$ increasing) between two consecutive critical points (i.e. for $j=i+1$) describe the RG flows (and the phase transitions) that occur in the QFT$_2$. 

Let us briefly remind how one can extract the information about the critical properties of such QFT$_2$ models from eq.(\ref{rg}) and the way the CFT$_2$ data is related to the asymptotic behaviour  of the NMG's domain wall solutions \cite{nmg}, or equivalently to the shape of the matter potential $V(\sigma)$.


\subsection {QFT$_2$ critical behaviour}

The zeros $\sigma^*_{A}$ of the $\beta$-function determine a set of \emph{critical points} in the coupling space, where the corresponding QFT$_2$ becomes conformal invariant and the phase transitions of second or infinite order take place. The nature of the observed changes in the behaviour of the thermodynamical (TD)  potentials and certain correlation functions at the neighbours of each critical point $\sigma_A^*$ does depend on the multiplicity $n_A$ of these zeros. In the case of simple zeros, we have $y(\sigma^*_{A}) = - d\beta / d\sigma |_{\sigma = \sigma^*_{A}} \neq 0$ and hence $\beta(\sigma) \approx -y(\sigma^*_{A})(\sigma- \sigma_A^*)$. The corresponding \emph{second order phase transitions}  are characterized by the scaling laws  and the critical exponents  of their TD potentials as for example $y_A = 1/\nu _A$, related to the singular part (s.p.) of  the reduced free energy per unit 2d volume $F^A_s = e^{2l}$ and to the correlation length $\xi_{A}=e^{-l}$ :
\begin{eqnarray}
 F^A_s(\sigma)\approx \left(\sigma- \sigma_A^*\right)^{\frac{2}{y_{A}}}, \quad\quad\quad \xi_A \approx(\sigma - \sigma_A^*)^{-\frac{1}{y_A}}  ,   
\label{sl}
\end{eqnarray} 
 at the neighbourhood of $\sigma_A^*$. Once the $\beta$-function (\ref{rg}) is given \footnote{Conjectured as in the case of the holographic RG or perturbatively calculated from the explicit form of the pCFT$_2$ action (\ref{eq28}).}, the  above ``near-critical forms" of $F^A_s(\sigma)$  and $\xi_A$  can be easily derived from the following RG equations:
\begin{eqnarray}
 \beta(\sigma)\frac{dF_s(\sigma)}{d\sigma} + 2F_s(\sigma)=0,\quad\quad \quad \beta(\sigma)\frac{d\xi(\sigma)}{d\sigma} =\xi(\sigma), \label{fs}
\end{eqnarray}
which determine the scaling properties of the TD potentials, etc. under infinitesimal RG transformations (see for example \cite{cardy}).

If one divides the  coupling space  $\s\in R$ into intervals $p_{k, \, k+1} = (\s_*^k, \s_*^{k+1})$ limited by vacua $\s_*$, then each interval will correspond to a different phase. Two such consecutive phases share the same UV critical point $\s_\uv^k$, where a second order phase transition, driven by a relevant operator $\Phi_\s$, may occur. The nature of this phase transition indeed depend on the properties of the neighbours, i.e. if $\s_*^{k \pm 1} = \s_\ir , \, \s_s , \, \infty$, which also determine  the features of the considered the QFT$_2$ - phase: massive, massless, etc. 
An efficient method for the analytic description of these QFT$_2$'s  phase transitions is given by the conformal perturbation theory $pCFT_{2}(\sigma^k_{\uv})$, based on the action (\ref{eq28}) and on the knowledge of the exact correlation functions of $\Phi_{\sigma}$, once the $CFT_{2}(\sigma^k_UV)$ is known and the relevant operator $\Phi_{\sigma}$ is appropriately chosen  \cite{x}.
 In the case of integrable perturbations of $\Phi_{13}$-type\footnote{These, for unitary models, are known to be the only consistent one coupling perturbations.} for Virasoro and Liouville (minimal) models (or of $\Phi_{adj}$-type for, say, $W_N$ m.m.s) \cite{x,fat,gms} the calculations involving conformal OPEs : 
 \begin{eqnarray}
 \Phi(1)\Phi(2)\approx I+C_{\Phi\Phi\Phi}\Phi(2)+...\label{ope}
 \end{eqnarray}
allow us to derive the $\beta$-function at first order in perturbation theory around the critical point: 
 \begin{eqnarray}
 \beta(\sigma)\approx -y_{13}(\sigma- \sigma_A^*)+C_{\Phi\Phi\Phi}(\sigma- \sigma_A^*)^2 +...\label{pertrg}
 \end{eqnarray}
It is well known that the phase structure of such pCFT$_2$ is of massless-to-massive $(\sigma^{\ir},\sigma^{\uv},\infty)$ type \cite{x}.

\subsection{On the NMG$_3$/QFT$_2$ correspondence}

We begin our short $NMG_3/pCFT_2$ dictionary by remembering one specific ``NMG feature" \cite{nmg} --- the existence of  two types of distinct \emph{critical} points: the usual \emph{type} (a) vacua, given by $W'(\sigma_{a}^*)=0$, and therefore representing the extrema of $W(\s)$; and the ``new'' vacua of \emph{type} (b), given by the real solutions of the equations $W^2(\sigma_{b}^*) = 2\epsilon m^2 / \kappa^2$, which exist only in the case when $\epsilon m^2 >0$. Both types of vacuum are extrema of the matter potential, $V'(\sigma_A^*)=0$, for
\begin{eqnarray}
\kappa^2V'(\sigma)=4W'\Big(1-\frac{W^2\kappa^2}{2\epsilon m^2}\Big)\omega(\sigma) ,
\end{eqnarray}
but there are others extrema of $V(\s)$, given by the real constant solutions of the algebraic equation:
\begin{eqnarray*}
\omega(\sigma^{*})=W''\Big(1-\frac{\kappa^2W^2}{2\epsilon m^2}\Big)-\frac{\kappa^2}{\epsilon m^2}(W')^2W-\epsilon\kappa^2W=0 ,
\end{eqnarray*}
which \emph{do not} represent (vacuum) solutions of the $I^{st}$ order eqs. (\ref{sis}). We will fix our attention, in what follows, in the vacua of type (a) and (b). As one can see from eqs.(\ref{eq4}), such vacua are defined by $\dot{\sigma}=0$ and $\dot{\varphi}=-2\epsilon \kappa W(\sigma_{A}^*)= {\mathrm{const}}$. It is then evident that they both present the geometry of an AdS$_3$ vacuum $(\sigma_A^*,\Lambda^A_{\eff})$ of the NMG model:
\begin{eqnarray*}
ds^2=dz^2+e^{-2\epsilon\sqrt{|\Lambda_{\eff}^A|}z}(dx^2-dt^2), \quad\quad A=a,b \; .
\end{eqnarray*}
As usually, the corresponding effective cosmological constants $\Lambda^A_{\eff}$ are realised as the vacuum values of  the 3d scalar curvature $R(\varphi)$, which, for the considered DW's and vacua solutions (\ref{intro1}), is given by
\begin{eqnarray}
R=-2\ddot{\varphi}-\frac{3}{2}\dot{\varphi}^2 = 8\epsilon(W')^2\left(1-\frac{\kappa^2W^2}{2\epsilon m^2}\right)-6\kappa^2W^2  \; ;  \label{curvature}
\end{eqnarray}
hence at a vacuum $\s^*_A$ we have $R_{vac} = -6\kappa^2W^2(\sigma_{A}^*) = 6\Lambda_{\eff}^A= - 6 / L_A^2$. Notice that the NMG vacua of $W(\sigma^*) = 0$ have the geometry of \emph{flat} Euclidean E$_3$ or Minkowski M$_3$ space.

The critical exponents also play a crucial part on the asymptotic behaviour of the matter field $\s(z)$. In the non-degenerated case we have
\begin{eqnarray}
\sigma(z)\stackrel{z\rightarrow\infty}{\approx}\sigma_{A}^* - \sigma_{A}^{0}e^{-
y_A\sqrt{|\Lambda^A_{\eff}|}z}, \quad
\Delta_A=2-y_A=1+\sqrt{1-\frac{m_{\sigma}^2(A)}{\Lambda_{\eff}^A}}, \quad \ m_{\sigma}^2=V''(\sigma_A^*) \; .   \label{asymp}
\end{eqnarray}
Thus $y_A\neq 0$ provide the boundary conditions (b.c.'s) for the corresponding DW's solutions of the NMG model (see ref.\cite{nmg}), as one can easily verify by considering the near-boundary/horizon approximation of eqs.(\ref{sis}): $\dot{\sigma}=-\epsilon \kappa \beta(\sigma)W(\sigma)\approx y_A (\sigma - \sigma_A^*)\epsilon \kappa W(\sigma_A^*)$, and taking into account the identification $\kappa W_A= - \epsilon / L_A$. 

As it is expected, the quantities characterizing the pCFT$_2$ present in the asymptotic limits of these QFT$_2$ can also be described by the geometric properties of the associated NMG-matter model, and written in terms of the superpotential. Let us first consider the critical exponents $y_{A}=y(\sigma_{A}^{*})=-\beta'(\sigma_{A}^{*})$ in the case when all the critical points have multiplicities $n_A=1$, i.e.  both  $\sigma_{a}^*$ and $\sigma_{b}^*$ are first order zeros of $\beta(\sigma)$, and $W(\sigma_{A}^*)\neq 0$\cite{nmg,oldholo}:
\begin{eqnarray}
y_{a}=y(\sigma_{a}^{*})=\frac{2\epsilon W''_{a}}{\kappa^2W_{a}}\Big(1-\frac{\kappa^2W_{a}^2}{2\epsilon m^2}\Big),\quad
y_b=y(\sigma_{b}^{*})=-\frac{4\epsilon(W'_{b})^2}{\kappa^2W_{b}^2},\quad \ W_{b}^2=\frac{2\epsilon m^2}{\kappa^2}  .    \label{sdim}
\end{eqnarray}

The structure constants $C^{A}_{\Phi\Phi\Phi}=\frac{1}{2}\beta^{''}(\sigma^*_{A})$ can be calculated from eqs.(\ref{rg}): 
\begin{eqnarray}
C^a_{\Phi\Phi\Phi}=- \frac{\epsilon W_a^{'''}}{\kappa^2 W_a}\bigg(1-\frac{W_a^2\kappa^2}{2\epsilon m^2}\bigg),\quad 
C^b_{\Phi\Phi\Phi}=\frac{2\epsilon}{\kappa^2}\Bigg(3\frac{W_b^{''}W_b^{'}}{W_b^2}- \frac{(W_b^{'})^3}{W_b^3}\Bigg) .   \label{str}
\end{eqnarray}
By definition, their $CFT_{2}(\sigma^k_{\uv})$ counterparts represent the ratio of the constants of 3-point and 2-point functions of the perturbing field $\Phi_{\sigma}$ \cite{bpz}. Finally, the Zamolodchikov's $C$-function can be written in terms of the superpotential as follows\cite{x}:
\begin{equation}
C(\sigma)= \frac{-3}{2G\kappa W(\sigma)}\bigg(1+\frac{\kappa^2W^2(\sigma)}{2\epsilon m^2}\bigg) \; . \label{cf}
\end{equation} 
This particular form is derived in refs. \cite{8,sinha,nmg} by the Brown-Henneaux  asymptotic method \cite{9}.
The central charge at a vacuum $c_A = C(\s_A^*)$ can therefore be evaluated when the superpotential is given.
It is important to note a specific feature of the NMG-induced 2d models, namely that the CFT$_2$'s  describing all the type (b) critical points have equal  central charges $c_{b}= 3L_{gr}/l_{pl}$ (with $L_{gr}^2= 1 / 2\epsilon m^2$)\footnote{corresponding to the lower  bound $\lambda_b=-1$ of the negative BHT-unitary window(\ref{bhtun})}, while the type (a) central charges $c_a = (3\epsilon L_a/2l_{pl})\left(1+ L_{gr}^2/L_a^2 \right)$ are parametrized by the corresponding critical values of the superpotential: $W^2(\sigma^*_a)= 1 / \kappa^2 L_a^2$.

\setcounter{equation}{0}
\section{Strong-weak coupling duality} \label{Duality}

Motivated by the eventual existence of holographic self-dual pCFT$_2$'s \footnote{representing few critical points and having massive and massless phases}, we address the problem about the properties of the pairs of the their dual 3d NMG-matter models and about the nature of the ''duality'' transformations relating their superpotentials.


\subsection{Pairs of dual NMG models}

Given a NMG model coupled to scalar field $\s$ of superpotential $W(\s)$ (\ref{acaoo}), we define its \textit{dual} as a specific NMG model, whose scalar field $\til \s$ and superpotential $\til W(\til\s)$  are fulfilling the following two conditions:
\begin{equation}
\bullet \;\;\;\; \p(\s) = \til\p(\til\s)\; , 
\;\;\;\;\;\;\;\;\;\;\;\;\;\;\;\;\;\;\;\;\;\; \bullet \;\;\;\; W(\s) = \frac{1}{\ka^2 L_{gr}^2 \til W(\til\s)}  \label{def dual}
\end{equation}
We also impose an additional requirement that all the critical points $\s_k$ and $\til\s_k$  of the pair of superpotentials, i.e. $W'(\s_k)=0=\til W'(\til\s_k)$ correspond to true AdS$_3$ vacua : $W(s_k)\neq 0$ and $\til W(\til\s_k)\neq 0$. It is natural to expect that the above pairs of duals $NMG_3$ models are mapped by the $AdS/CFT$ correspondence rules in certain pairs of duals (or self-dual) $CFT_2$'s models. The particular form of the  NMG's matter superpotentials transformations (\ref{def dual}) is chosen in the way that  the  coupling space duality transformations between the corresponding pairs of duals $CFT_2$'s, induced by eqs.(\ref{def dual}), preserve the form of the central charges at the critical points, the form of the central function (\ref{cf}) and of the corresponding s.p. of their free energy.

The first requirement, i.e. the invariance of the  scale factor of the NMG's domain walls, ensures the desired invariance of singular part of the reduced free energy $F^A_s (\s)= e^{-\p(\s)}$ of theirs duals pCFT$_2$ models. It is equivalent to the condition  $l(\s) = \til l(\til\s)$ of the invariance of the QFT$_2$ scales under such transformation. We next recall that the central charge associated with a vacuum $\s_A$  has the form:
\be
c_A = \frac{3 \epsilon L_A}{2l_{pl}} \, \left[ 1 + \frac{L_{gr}^2}{L_A^2} \right] , \label{central charge duality}
\ee
where $L_A = \pm 1 / \ka W(\s_A)$ -- with the sign chosen in order to make $L_A$ positive --, and also that $L_{gr}^2 =  1/2m^2 \epsilon > 0$ denotes the radius of the type (b) vacua (cf. eq. (\ref{sis})).  Then the  second condition in (\ref{def dual}) implies 
\begin{equation}
\til L_A = \frac{L_{gr}^2}{L_A} .  \label{dual scales}
\end{equation}
Therefore the above NMG's superpotential transformation ensures the invariance of the $CFT_2$'s central charges\footnote{and of the corresponding pCFT$_2$'s  C-function (\ref{cf}) as well} and it has the same form as the well known central charges duality properties of the Liouville and minimal models, namely $c(L_A) = c(L_{gr}^2/L_A)$. Notice that
this is a direct consequence of the curvature quadratic terms in the action (\ref{acaoo}), which generates the specific form of the central charge (\ref{central charge duality}), which \emph{is not present} in EH gravity. The transformation $\s \to \til\s$ maps the AdS$_3$ spaces of large radii (and small cosmological constants) to certain ''dual'' AdS$_3$ spaces of small radii (and large cosmological constants)\footnote{Note that if the types (a) and (b) vacua coincide, then all the scales remain invariant: $L_a = L_{gr} = \til L_a$.}, but the corresponding dual $CFT_2$'s share \emph{equal} central charges.

An important (implicit) element of the above introduced  concept of pairs of duals NMG's (and corresponding pairs of duals pCFT$_2$'s) is that the  mapping is always between the vacua of the same kind, i.e. $\s_a \to \til\s_a$ and $\s_b \to \til\s_b$. This must be proved however. We first note that according to the condition $\p (\s) = \til\p (\til\s)$, the transformation of the $beta$-funtion $\beta(\s) = - d\s/d l$, with $l = - \p/2$ is given by:
\be
\beta(\s) = \frac{d \s}{d \til\s} \til\beta(\til\s) . \label{dual beta}
\ee
The next step is to calculate the derivative $d \s / d \til\s$ in terms of the corresponding  superpotentials, by  substituting the explicit form (\ref{rg}) of the both $\beta$-functions into eq.(\ref{dual beta}), and then taking into account eqs.(\ref{def dual}) to eliminate $\til W$ : 
\be
 d \til\s/d \s = - 1/\ka L_{gr} \, W(\s). \label{dutrans}
\ee
Due to the additional requirements $W(\s_k)\neq 0$ and $\til W(\til\s_k)\neq 0$ it is not singular at the critical points, and therefore the  zeros of $\beta(\s)$ are also zeros of $\til\beta(\til\s)$ and vice-versa. Hence the vacua of one theory are also vacua of its dual, and the transformation (\ref{def dual}) maps vacua into vacua. As a consequence we find the explicit form of the NMG's scalar fields duality transformation as follows: 
\be
\til\s(\s) =  - \frac{1}{\ka L_{gr}} \, \int^\s \! \frac{d x}{W(x)} + {\mathrm{constant}} . \label{int s til}
\ee
The above properties confirm the fact that the type (a) NMG -vacua are mapped into the type (a) vacua of the dual NMG model $\s_a \rightarrow \til\s_a$, as one can see from the identity $d \til W(\til\s) / d \til\s = (\ka L_{gr} \, W(\s))^{-1} d W(\s) /d \s$. The type (b) vacua remain invariant under the duality transformation, since their  defining equation $1 - \ka^2 L_{gr}^2 W^2(\s) = 0$ is mapped by (\ref{def dual}) into itself.

Taking into account the explicit form of the I-st order eqs.(\ref{sis}) for the pairs of NMG dual models, it is not difficult  to derive the relation between the dual ``radial" coordinates $\til z(z)$:
\be
\til z(z) = \ka^2 L_{gr}^2 \int^z \! d x \; W^2(x) + {\mathrm{constant}} ,
\ee
or, in terms of $\s$ we get
\be
\til z(\s) =  \frac{\ka^2 L_{gr}^2}{2} \int^\s \! \; \frac{ W^2(x) \, dx}{W'(x) \left[ 1 - \ka^2 L_{gr}^2 W^2(x) \right]} + {\mathrm{constant}}.
\ee
It remains to demonstrate one of the most important properties of the duality transformations (\ref{def dual}): namely, that they keep invariant the critical exponents $y_A$, $A = a, b$ given by (\ref{sdim}). Starting by their definitions $y_A = d \beta / d \s \mid_{\s_A}$ at the corresponding critical points $\s_A$, and further by  using eqs. (\ref{cf}), and the fact that the vacua are the zeros of these $\beta$-functions, we find
$$ y_A = - \frac{d}{d \s} \beta (\s) \mid_{\s_A} = - \left\{ \frac{d \til\s}{d\s} \, \frac{d}{d \til\s} \left[ \frac{d\s}{d\til\s} \, \til\beta(\til\s) \right] \right\}_{\til\s_A} = - \left\{ \frac{d \til\s}{d\s} \frac{d\s}{d\til\s} \, \frac{d}{d \til\s} \til\beta(\til\s) \right\}_{\til\s_A} = - \frac{d}{d \til\s} \til\beta(\til\s) \mid_{\til\s_A} $$
Thus we can conclude that indeed $y_A = \til y_A$.

Let us summarize the main features of the duality transformations (\ref{def dual}) between two specific NMG -matter models, whose superpotentials are  ''inversely proportional'': their matter potentials are different, but  they do have  equal number of vacua such that the pairs of type (a) dual vacua are representing  $AdS_3$ spaces of different radii that are inversely proportional to each other and their  type (b) vacua are coinciding. The most relevant characteristics of the corresponding pairs of dual $CFT_2$'s models (and of the pairs of pCFT$_2$'s as well) are: (1) they have different holographic $\beta$-functions, whose type (a) critical points  have different values (one in the weak-coupling another in the strong coupling regions), but still identical central charges and central functions; (2) the critical exponents $y_A = \til y_A$ remains invariant under such duality transformations and (3) their s.p. free energies are identical by construction. It remains to answer the important question concerning the explicit construction of relatively simple and physically interesting pairs of such dual NMG models and to describe the  nature phase transitions and of the different phases of the corresponding pairs of duals pCFT$_2$'s, whose exact holographic $\beta$-functions are related by the eqs.(\ref{dual beta}).


\subsection{Examples of dual and self-dual NMG models}	\label{Examples of dual and self-dual}

In order to illustrate how the concepts  of NMG duality transformations (\ref{def dual})  introduced above can be realized in practice, we consider few representative simple examples of pairs of NMG dual models. An important problem addressed in this subsection  concerns one particular class of duality transformations $\s=\s(\til\s)$, that together with the definitions (\ref{def dual}) and (\ref{int s til}) satisfy the new ''self-duality'' condition: namely, when substituted in the second of the eqs.(\ref{def dual}) to give rise of a very special self-dual superpotentials: 
\be
\bullet \;\; \text{self-duality}:\;\; W(g_k,\s)=\til W(g_k,\til\s),\;\;\;\;\;\;\;\;\;\bullet\;\;\text{partial self-duality}:\;\; W(g_k,\s)=\til W(\til g_k,\til\s),
\ee
where the parameters $g_k$ and $\til g_k$ determine the coupling constants and the masses in the corresponding $NMG_3$ matter potentials $V(g_k,\s)$ and $\til V(\til g_k,\til\s)$. In both cases the shapes of the pairs of duals NMG  superpotentals are coinciding, but in the second case the particular ``partial self-duality'' transformations are  mapping  the NMG-matter couplings $\til g_k=\til g_k(g_k)$ as well. The particular examples analysed in this section are all chosen to provide a kind of ''strong-to-weak couplings'' duality transformations $\s=\s(\til\s)$ between the corresponding pairs of dual pCFT$_2$'s.


\subsubsection{Self-duality}  \label{self-duality}

Consider the following quadratic superpotential:
\be
W(\s) = B \s^2 \; , \;\; B > 0 .\label{linear}
\ee
We assume that there exist at least one (b) vacuum, i.e. $m^2\epsilon > 0$, which is the fixed point of the transformation (\ref{def dual}). Because of the $Z_2$ symmetry of the superpotential, we can consider the $\s > 0$ only. There is no type (a) vacuum for such superpotentials: the vacuum at $\s_M = 0$ is of zero cosmological constant, i.e. it represents a Minkowski vacuum. The exact form of the scale factor is easily derived by solving the corresponding I-st order system (\ref{sis}) and it determines a particular asymptotically AdS$_3$ ( or H$_3$ in the euclidean case) geometry with a naked singularity at $\s \rightarrow \infty$ \cite{nmg,oldholo}. The eqs.(\ref{def dual}) and (\ref{int s til}) applied  for the linear $W$ (\ref{linear})  provides the explicit form of the NMG duality transformation:
\be
\til \s = \frac{1}{\ka L_{gr} B \, \s} ,\;\;\;\;\;\;\;\;\;\;\; \;    \til W(\til\s) = 1 / \ka^2 L_{gr}^2 B \s^2 = B \til\s^2, \label{sdu}
\ee
where the constant of integration has been chosen to be zero. Therefore the dual superpotential has exactly the same shape of the original one and coinciding parameters $B=\til B$, that determine the coupling constants in the corresponding matter potentials $V(\s)$ and $\til V(\til\s)$. The critical points, however, are `` interchanged'' in the dual model: the original Minkowski vacuum is mapped into the dual naked singularity and the original naked singularity is mapped into the dual Minkowski vacuum.

\subsubsection{Partial self-duality} \label{partial self-duality}

The simplest example of partially self-dual NMG -models is given by the following hyperbolic superpotential: 
\be
W(\s) = B \, \sinh(D \s) \, \;\;\quad  B > 0.
\ee
It does not lead to physically interesting self-dual pCFT$_2$, due to the fact that, similarly to the linear superpotential model considered in the beginning of this section, it has only one type (b) vacuum at $\s_b = D^{-1} \sinh^{-1} \{ (B \ka L_{gr})^{-1} \}$, a naked singularities at $\s \rightarrow \pm \infty$ and no one type (a) vacua \footnote{Although there is no problem with the geometry, the $\beta$-function diverges at $\s = 0$, so the holographic description is not well defined in this point.}. The explicit form of the corresponding duality transformation (\ref{int s til}) can be found by simple integration:
\be
\cosh( D \s ) =  \coth \left(\ka L_{gr} B D \til\s \right) , \label{sh s e til}
\ee
By substituting it in the defining equation (\ref{def dual}), we deduce the following form of the dual superpotential:
  $$\til W(\til\s) = \til B \sinh (-\til D \til\s) , \;\; {\mathrm{with}} \;\;  \til B = \frac{1}{\ka^2 L_{gr}^2 B} , \;\; \til D = \ka L_{gr} BD. $$ 
Therefore  the original duality transformation (\ref{def dual}) in the case of the hyperbolic superpotential leaves invariant its shape, but it is changing its parameters. The true self-duality is achieved for a specific ''critical'' value of $B$, namely $B = 1 / \ka L_{gr}$. 

\vspace{0.5cm}

We next consider another example of partially self-dual superpotential:
\be
W(\s) = \left[ B (\s - \s_a)^2 + D \right]^{3/2} , \;\; D > 0 . \label{W dmn}
\ee
that give rise to an interesting strong-weak coupling self-dual  pCFT$_2$, representing dual massive and massless phases and also few self-duals $CFT_2$'s describing its (a) and (b) type vacua. The type (a) vacuum is placed at the critical value $\s = \s_a$ with $\kappa L_a = D^{-3/2}$ an its  type (b) vacua at 
\be
 \s_b^{\pm} = \s_a \pm \sqrt{\frac{1}{B (\ka L_a)^{2/3}} \left[ \left(\frac{L_a}{L_{gr}} \right)^{2/3} - 1 \right]} . \label{sigma b gmn}
\ee
Their number depends on how many real values $\s_b^\pm \in \mathbb{R}$  can take. Thus, the existence and the number of the type (b) vacua  is determined by the sign of $B$ and on the values of  the ratio $L_a/L_{gr}$. Notice that, if $B < 0$, there are Minkowski  vacua at 
\be
\s_M^\pm = \s_a \pm \sqrt{\frac{D}{|B|}} ,
\ee
allowing the relation (\ref{sigma b gmn}) to be written as
\be
\frac{(\s_b - \s_a)^2}{(\s_M - \s_a)^2} = 1 - \left( \frac{L_a}{L_{gr}} \right)^{3/2} ,
\ee
which is valid for $B < 0$ only. Since for $B < 0$ we have $L_a < L_{gr}$, the relation above shows that $0 < (\s_b - \s_a) / (\s_M - \s_a) < 1$, i.e. the Minkowski vacua are farther from $\s_a$ than the type (b) vacua.

We complete our description of the vacua structure of the NMG model with superpotential (\ref{W  dmn}) by listing  all the possible different sets of allowed vacua, depending on the signs and the values of the parameters of this superpotential (see fig.1). In all the cases there exists one type (a) vacuum. With regard to the other vacua, we have:

\begin{itemize}
\item (I) : $L_a > L_{gr}$
\subitem I.a . $B > 0$ : There are vacua of type (b);
\subitem I.b . $B < 0$ : There are Minkowski vacua;

\item (II) : $L_a < L_{gr}$
\subitem II.a . $B > 0$ : There are no Minkowski nor type (b) vacua;
\subitem II.b . $B < 0$ : There are both Minkowski and type (b) vacua;

\item (III) (critical case) : $L_a = L_{gr}$
\subitem III.a . $B > 0$ : The only vacuum is $\s_{a} = \s_b$;
\subitem III.b . $B < 0$ : There are Minkowski vacua as well as $\s_a = \s_b$.
\end{itemize}

The explicit form of the  duality transformation (\ref{int s til}) specific for the considered superpotential (\ref{W dmn}) is given by: 
\be
\til\s - \til\s_a = \frac{L_a}{L_{gr}} \, \frac{\s - \s_a}{\sqrt{1 + \frac{B}{D}(\s - \s_a)^2}} , \label{til sigma gmn}
\ee
where the (arbitrary) integration constant is denoted by  $\til\s_a$. It determines the position of the (a) type vacua dual to the original type (a)  one, i.e. we have $\s_a \rightarrow \til\s_a$ under the duality transformation (\ref{til sigma gmn}). This arbitrariness can (and will) be used to fix one of the (b) vacua $\s^\pm_b$ as a fixed point of the duality transformation. 

Substituting eq. (\ref{til sigma gmn}) into (\ref{def dual}), one derives the form of the dual superpotential
\be
\til W(\til\s) =  \left[ \til B(\til\s - \til\s_a)^2 + \til D \right]^{3/2} ,\quad \til B = - ( L_{gr} / L_a )^{2/3} \, B \; ,\;\;\quad \til D = \frac{1}{(\ka L_{gr})^{4/3} D} . \label{dual param. gmn}
\ee
The last equation for $\til D$ was to be expected, since it reflects only the fact that $L_a = L_{gr}^2 / \til L_a$ (recall that $L_a = (\ka D)^{-3/2}$). The difference of sign between the dual superpotentials is not important\footnote{The global sign of the superpotential (or it's dual) is relevant only in the identification $W(\s_A) = \pm 1 / L_A$, where the sign must be chosen in order to make $L_A$ positive.}, and $\til W(\til\s)$ has the same vacua structure as it's dual, which is described by cases $(Ia)$, etc. above -- but now with the ``tilde'' quantities $\til B$, $\til L_a$, etc. Since the transformation (\ref{dual param. gmn}) changes the sign of $B$, i.e. $B/\til B < 0$, we establish  the duality equivalence between the following models :
$$ {\mathrm{(I.a)}} \Leftrightarrow {\mathrm{(II.b)}} \; ; \;\;\;  {\mathrm{(I.b)}} \Leftrightarrow {\mathrm{(II.a)}} \; ;  \;\;\;  {\mathrm{(III.a)}} \Leftrightarrow {\mathrm{(III.b)}}$$
as one can see on fig.2. The most interesting case is the first one, so we will analyse it in more detail. It corresponds to $L_a > L_{gr}$ and $B > 0$, thus $\til L_a < L_{gr}$ and $\til B < 0$.

\begin{figure}[ht]
    \centering
    \includegraphics[scale=0.6]{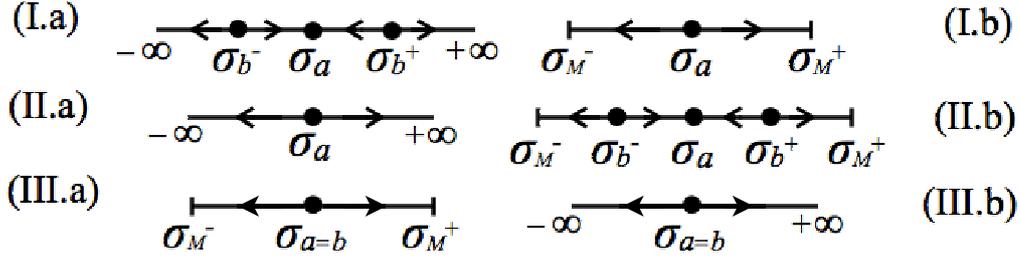}
    \begin{quotation}
    \caption[ed]{\small Graphic representation of the admissible vacua stricture and of the expected RG flows for $\epsilon=-1$ and $m^2<0$.}
     \label{fig:1}
     \end{quotation}
\end{figure}

The vacua structure compiled in the cases (I) to (III) above is not complete without the information about the stability (UV  versus IR) of the corresponding vacua, according to the sign of the critical exponents $y_A$ given by (\ref{sdim}). We have
\br
 y_{a} = - 6 B \frac{L_a^{2/3}}{\ka^{4/3}} \, \left[ 1 - \left( \frac{L_{gr}}{L_a} \right)^2 \right] ; \;\;\; y_b =  4 B \frac{L_{gr}^{2/3}}{\ka^2}  \left[ 1 - \left( \frac{L_{gr}}{L_a} \right)^{2/3} \right] .
\er
and therefore in the cases (I.a) and (II.b) where $y_a < 0$ - the type (a) vacuum is an IR critical point. The type (b) vacuum has $y_b > 0$ and hence it corresponds to an UV critical point. In the cases (I.b) and (II.a) the sign of $y_a$ is reversed, i.e. $y_a >0$ and now  the type (a) vacua are representing the UV critical points. 

The type (b) vacua $\s^\pm_b$ are mapped into the type (b) vacua $\til \s_B^\pm$ of the dual theory through eq. (\ref{til sigma gmn}):
\be
\til\s_b^\pm - \til\s_a = \left( \frac{L_a}{L_{gr}} \right)^{2/3} (\s_b^\pm - \s_a) . \label{til s b dmn}
\ee
As said before, the constant of integration $\til\s_a$ can be chosen in order to set one of the type (b) vacua as a fixed point of the duality transformation, namely $\s_b^- = \til\s_b^-$. Thus we must have
\be
\til\s_a = \left( \frac{L_a}{L_{gr}} \right)^{2/3} \s_a - \left[ \left( \frac{L_a}{L_{gr}} \right)^{2/3} - 1 \right] \s_b^- .
\ee
On the other hand, the constant $\s_a$ is also arbitrary, since it can be changed by a translation of $\s$. Hence we can further adjust it in order to put the fixed point $\s_b^-$ at the origin. By taking 
\be
\s_a = \sqrt{ \frac{1}{B (\ka L_a)^{2/3}} \, \left[ \left( \frac{L_a}{L_{gr}} \right)^{2/3} - 1 \right] } , \label{sigma a fixed}
\ee
we get $\s_b^- = 0$, and also that $\s_b^+ = 2 \s_a$ (cf. eq. (\ref{sigma b gmn})). An important consequence of this choice is that the values of the corresponding critical couplings $\til\s_a$ of the dual model
\be
\til\s_a =  \left( \frac{L_a}{L_{gr}} \right)^{2/3} \s_a ,
\ee
 are  greater then $\s_a$, i.e.  $\til\s_a > \s_a$ in the cases  (I.a) and (I.b), when we have $L_a > L_{gr}$.  Therefore in the asymptotic regime of very large scales $L_a >> L_{gr}$, we realize that the weak coupling critical point $\s_a - \s_b^- = \s_a << 1$, is mapped to the strong coupling regime of the dual model since now we have that $\til\s_a - \til\s_b^- = \til\s_a >> 1$.

\begin{figure}[ht]
    \centering
    \includegraphics[scale=0.45]{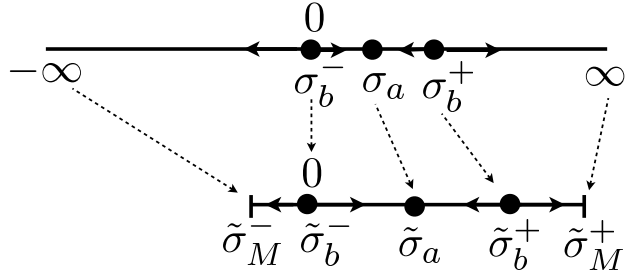}
    \begin{quotation}
    \caption[ed]{\small Symbolic diagram demonstrating the duality between different phases in the case $L_a>L_b$ and $B>0$.}
     \label{fig:2}
     \end{quotation}
\end{figure}

In order to find the scale factor, we integrate the $\beta$-function equation, obtaining
$$ \p(\s) =  - \frac{\ka^2}{3 \e B} \int \! \frac{B(\s-\s_a)^2 + D}{(\s - \s_a) \left\{ 1 - \ka^2 L_{gr}^2 \left[ B (\s - \s_a)^2 + D \right]^3 \right\}} \, d\s + {\mathrm{const}} .$$
With the substitution $g(\s) = B(\s-\s_a)^2 + D$, we get
\br
& \p(\s) = \frac{1}{6 \epsilon B L_{gr}^2} \sum_{i=0}^3 A_i \log(g - g_i) + \p_{\infty}, \nonumber\\ 
& e^{\p(\s)} = e^{\p_\infty} \,  \prod_{i=0}^3 |g - g_i|^{x_i},\quad x_i = A_i / 6 \epsilon B L_{gr}^2\nonumber
\er  
where 
\br 
&g_1 = \ka^{-2/3} L_{gr}^{-2/3} ,\quad g_2 = g_3^* = -(\ka L_{gr})^{-2/3} \left(\frac{1 + i \sqrt{3}}{2} \right),\quad A_0 \equiv A_a = \frac{- \ka^{4/3} L_{gr}^2L_a^{-2/3}}{ 1- (L_{gr} /L_a)^{2} },\nonumber\\
&g_0 \equiv g_a = D, \quad  A_1 = \frac{ \ka^{4/3} L_{gr}^{4/3}}{3 \left[ 1- (L_{gr}/L_a)^{2/3} \right]}, \quad A_2 =  A_3^* = \frac{  \ka^{4/3} L_{gr}^{4/3} \, (i -  \sqrt{3})}{3 \left[2 i +  \left( i +  \sqrt{3} \right) (L_{gr}/L_a)^{2/3} \right]}. \label{A23}
\er
Notice that although both the exponents $x_{2,3}$ and the roots $g_{2,3}$ are complex, the last two terms of the product above are real, and so is the expression for the scale factor. Also notice that $\sum x_i = 0$, hence if $\s \to \infty$ we have $e^\p \to e^{\p_\infty}$, allowing for the possibility of a naked singularity. This property also allow us to rewrite the scale factor more explicitly as
\br
e^{\p(\s)} = e^{\p_{\infty}} \, (\s - \s_a)^{2 x_a} \, (\s - \s_b^+)^{x_1} (\s - \s_b^-)^{x_1} \prod_{i=2}^3 \left[ (\s - \s_a)^2 + (D - g_i) / B \right]^{x_i} , \label{scale factor dmn}
\er
where $\s_b^- = 0$ and $\s_b^+ = 2 \s_a$, with the constant $\s_a$ being given by (\ref{sigma a fixed}). It is now evident the singular behaviour of the scale factor (and hence of the correlation length) near to the vacua (i.e. the critical points). 

It is worthwhile to mention that the particular (degenerate) case $D=0$ leads to superpotential: 
\begin{eqnarray}
W(\sigma)=E(\sigma-\sigma_M)^3,\quad\quad  \ E>0,
\end{eqnarray}
which has different vacua structure and in fact it is \emph{not any more partially self-dual}. The duality transformation in this case has the form:
\begin{eqnarray}
\sigma-\sigma_M=\sqrt{\frac{1}{2L_{gr}E}}\frac{1}{\sqrt{\tilde\sigma_M-\tilde\sigma}}, \ \quad\quad \tilde\sigma_M>\tilde\sigma,\label{aaaa}
\end{eqnarray}
which allows us to deduce the explicit form of the corresponding dual superpotential:
\begin{eqnarray}
\tilde W(\tilde\sigma)=\tilde E\left(\tilde\sigma_M-\tilde\sigma\right)^{3/2},\quad \;\;\;\;\;\;\tilde E=2^{3/2}\left(\frac{E}{L_{gr}}\right)^{1/3},\nonumber
\end{eqnarray}
Evidently we have an example of dual transformation that is \emph{not preserving} neither the shape of the superpotential nor the values of its parameters.

Let us also mention that  the examples  we have studied in this subsection, do not exhaust all the possible partially self-dual superpotentials of one or two type (a) vacua. Another physically interesting example of pairs of NMG's is given by the following periodic superpotential:
\be
W(\s) = B \left[ D - \cos (\a \s) \right] , \quad\quad B < 0 , \label{W cos} 
\ee
whose vacua structure -of \emph{two} type (a) vacua-, its duality properties and also certain features of the phases  of its dual $pCFT_2$ are described in the Appendix below. 

The construction of examples of self-duals and partially self-dual NMG's based on  superpotentials having more then \emph{two type (a) non-degenerate} critical points and the explicit forms (\ref{til sigma gmn}) of the corresponding duality transformations, involves relatively big number of W-parameters (i.e. the matter fields couplings $g_k$ as B,D etc.). It represents a  rather complicated open problem and  requires better understanding of the group properties of the couplings $\s$-transformations and of the group structure behind our definition of the 
partial self-duality, as well as further investigations of the group-theoretical nature of the W-parameters $\til g_k=\til g_k(g_k)$ transformations, see (\ref{dual param. gmn}).

\subsection{Unitary consistency of duality}

An important test for the physical consistency of the pairs of dual vacua, i.e. pair of AdS$_3$'s of dual radii $L_a$ and $\til L_a=L^2_{gr}/L_a$, is the verification of whether and under what conditions (if any) they both belong to the same BHT negative unitary window (\ref{bhtun}).
Let us first briefly remind the content of the BHT unitarity conditions for the NMG models \cite{1,more}. Remember that the negative value of $m^2_{\sigma}(A)$ for scalar fields (tachyons) in AdS$_3$ backgrounds, which appears in the dimensions of the relevant operators,  do not cause problems when the Breitenlohner-Freedman (BF) condition \cite{BF},
\begin{eqnarray}
\Lambda_{\eff}^A\le m_{\sigma}^2(A), \label{A3} 
\end{eqnarray}
is satisfied. The unitarity of the purely gravitational sector of the NMG model (\ref{acao}) requires that the following Bergshoeff-Hohm-Townsend (BHT) conditions \cite{1,more}:
\begin{eqnarray}
m^2\left(\Lambda_{\eff}^A-2\epsilon m^2\right)>0,\quad\quad
\Lambda_{\eff}^A\le M_{gr}^2(A)=-\epsilon m^2+\frac{1}{2}\Lambda_{\eff}^A, \label{bht}
\end{eqnarray}
take place. Taking into account that for the each vacua $\s_A$ the BHT -parameter $\la_A$ can be realized as follows:  
\begin{eqnarray}   
&&\la_a=\la(\s^*_a)=\frac{\ka^2}{2m^2}V(\s^*_a)=\frac{L^2_{gr}}{L^2_a}\big(\frac{L^2_{gr}}{L^2_a}-2\big), \label{launit}
\end{eqnarray}   
which for the type $(b)$ vacuum, i.e. $W_{\pm}^2=\frac{2\epsilon m^2}{\ka^2}$, reproduces the lower bound $\la=-1$ of BHT-condition (\ref{bhtun}).

In order to derive the unitarity  restrictions on the generic type $(a)$ vacuum we introduce the following notation: $q=\frac{\Lambda_{eff}^{a}}{\Lambda_{eff}^{b}}=\frac{\ka^2 W_{*}^2}{2\epsilon m^2}$. Then we have $\la_{(a,b)}=q(q-2)\equiv \la(q)$,
which makes evident that $\la(q)=\la(2-q)$. Therefore the $\la_a$ values for which the unitarity condition (\ref{bhtun})) is satisfied are imposing restrictions on the allowed $L_a$ values:
\begin{eqnarray}
0\leq \frac{L^2_{gr}}{L^2_a} \le 2, \quad\quad \epsilon=-1, \quad\quad  m^2<0 \label{bhtlaa}
\end{eqnarray}
and consequently on  the central charges (\ref{ch}) of the corresponding CFT's. The type (b) NMG vacua are known to be always unitary \cite{nmg} of $\la=-1$ and whether it represents  UV or IR critical points of the dual pCFT$_2$ depends on the sign factor only: UV for $\epsilon=-1$, since we have $y_b>0$, and IR for $\epsilon=1$. The properties of the type (a) critical points (UV or IR) do depend on both the sign of $\epsilon$ and the particular form of the matter superpotential, as one can see from eq.(\ref{sdim}). The unitarity of the NMG-matter model is still an open problem, and it requires further analysis of the linear fluctuations around the DW's relating, say, two unitary BHT-vacua from the negative \emph{BHT-unitary window}: $-1\leq \la<0, \epsilon=-1$, $m^2<0$. We are however obliged to require that at least all the NMG-matter model's vacua are BHT-unitary. 

In the context of the NMG duality transformations, when applied for the critical points $\til L_a = \frac{L_{gr}^2}{L_a}$, we impose an additional condition, namely  that the  ''dual'' scales $\til L_a$ and $L_a$  are both belonging to the same (negative) BHT - unitary window(\ref{bhtun}). Taking into account the eqs.(\ref{bhtlaa}) and (\ref{dual scales}) we conclude that the NMG duality (\ref{def dual}) is compatible with the NMG unitarity only when the following conditions are fulfilled : 
\begin{eqnarray}
   \frac{L_{gr}}{\sqrt{2}} \leq \til L_a \leq L_{gr} \leq L_a \leq L_{gr} \sqrt{2} \label{unidual}
\end{eqnarray}
Hence when $L_a$ and its dual scale $\til L_a$  both belong to certain finite interval of values $(L_{gr}/\sqrt{2}, L_{gr} \sqrt{2})$ they describe dual pairs of unitary NMG's vacua.

\subsection{On the group properties of partial self-duality}

One of the  main features of the strong-weak coupling duality is that in the self-dual 2-,3- and 4-dimensional (supersymmetric) QFT's, it is always realized as an inversion transformation and more generally as fractional linear transformations belonging to certain (discrete)\footnote{i.e. of $SL(2,Z)$ as for example in the cases of models having discrete spectrum of energies or/and charges- electric and magnetic etc.} 
 subgroups of $SL(2,C)$ \cite{seiberg,seibit,cardy-itz}. It is therefore important to verify whether these well known properties of the QFT's duality (or certain limits of them) take place in the particular examples of $pCFT_2$'s  duals to  (pairs of) NMG models with appropriately chosen superpotentials (\ref{def dual}). The question  about the gravitational $d=3$ NMG meaning of the $d=2$ conditions of  strong-weak coupling duality symmetries in the considered two dimensional $pCFT$'s is also addressed.

 Let us remind that the requirements on the NMG's superpotentials (\ref{def dual}), that select holographic self-dual $pCFT_2$'s, have been introduced  by extending the ''critical'' duality transformation $\til L_A = \frac{L_{gr}^2}{L_A}$ (at each critical point $\s^*_A$) to its ''off-critical'' equivalent (\ref{def dual}). As a consequence we have deduced the explicit form (\ref{int s til}) of the corresponding coupling's $\til \s =\til \s (\s)$ transformations that are keeping invariant the central charges, central functions, s.p. of the reduced free energy, but in principal they are changing the form of the exact holographic $\beta$-functions, according to eqs.(\ref{dual beta}). Notice that the $L_A$'s transformation (and the W's as well) represents a particular $G_L\in GL(2,R)$ transformation, i.e.
\be  
\til L_A=\frac{aL_A+b}{cL_A+d},\quad\quad G=\left( \begin{array}{cc}
a & b \\
c & d
\end{array} \right),\quad\quad G_L=\left( \begin{array}{cc}
0 & L^2_{gr} \\
1 & 0
\end{array} \right),\quad \quad G^{-1}_L=\left( \begin{array}{cc}
0 & 1 \\
\frac{1}{L^2_{gr}} & 0
\end{array} \right) \label{fraclin}
\ee 
By introducing their dimensionless counterparts, say $l_A=L_A/L_{gr}$ we indeed recover the well known standard large-small radii $Z_2$ inversion transformation:
\be 
\til l_A= 1/l_A, \quad\quad\quad \text{i.e.} \quad G_I=\left( \begin{array}{cc}
0 & 1 \\
1 & 0
\end{array} \right)=G^{-1}_I,
\ee  
such that the large $l_A\gg 1$ (i.e. $L_A\gg L_{gr}$) are mapped to the very small $\til l_A\ll 1$ ones (in the $L_{gr}$ units of length).

We next consider the problem of the similarities and the differences between  the group properties of the particular strong-weak coupling $pCFT_2$'s duality transformations  (\ref{sdu}) and (\ref{til sigma gmn}), present in the specific -one type (a) vacua-  examples of self-dual(SD)  and partially self-dual(PSD) models, studied in Sect.3.2.

\textit{Self-dual models.} The corresponding SD coupling's transformation is almost identical of the $L_A$'s ones: 
\be 
\til \s=\frac{\s^2_+}{\s} ,\quad\quad \sigma_+ = 1 / \sqrt{\ka L_{gr}B} \quad\quad \text{with} \quad\quad  \s_+ = \til \s_+ ,\label{sdtr}
\ee 
which takes the standard inversion form  $\til u_{sd}=1/ u_{sd}$ for the rescaled coupling $u_{sd}=\frac{\s}{\s_+}$. Notice that the strong couplings $\s\gg \s_+$ are mapped to the weak ones $\til \s\ll \s_+$. An important feature of this self-dual model is that the above SD transformations are leaving invariant \footnote{together with the free energy, central function and the anomalous dimensions} the RG equation:
\begin{eqnarray}
\frac{du_{sd}}{dl} = \frac{4\epsilon}{\kappa^2 u_{sd}}\big(1-u^4_{sd}\big)=-\beta_{sd}(u_{sd}),    \label{rgsd}
\end{eqnarray}
and the form of the corresponding exact $\beta-$function, i.e. we have  $\beta_{sd}(u)=\beta_{sd}(\til u)$ as well. It is important to mention that this $\beta_{sd}$-invariance property is indeed consistent with the general covariance requirement (\ref{dual beta}). It reflects a very particular form of our SD superpotential and related to it $\beta_{sd}$.

\textit{Partially self-dual models.} Let us consider the PSD  transformation (\ref{til sigma gmn}) for the square of the coupling $u=(\s-\s_a)^2$, i.e.\footnote{ we are simultaneously rescaling the B and D parameters in the way that the equivalent rescaling of the superpotential $w=\ka L_{gr} W$ leads to the standard inversion form of the duality condition (\ref{def dual}): $\til w(\til \s)=1/w(\s)$.}:  
\be 
 \tilde u = \left(\frac{1}{d^2}\right) \frac{u}{d+bu} , \quad\quad u>0, \quad d=(\ka L_{gr})^{2/3}D, \quad\quad b =(\ka L_{gr})^{2/3}B,\quad d>0, \quad b>0 .\label{psdconf}
\ee 
It is then evident that it represents a \emph{two parameters subgroup} of the (general) fractional linear transformations $G_{psd}(d,b)\in GL(2,R)$:  
\begin{eqnarray}
& G_{psd}=\left( \begin{array}{cc}
1/d^2 & 0 \\
b &  d
\end{array} \right)=\left( \begin{array}{cc}
1/d^2 & 0 \\
0 & d
\end{array} \right)\left( \begin{array}{cc}
1 & 0  \\
b/d & 1
\end{array} \right) ; \; G^{-1}_{psd}=\left( \begin{array}{cc}
d^2 & 0 \\
-bd & 1/d
\end{array} \right) =\left( \begin{array}{cc}
1/\til d^2 & 0 \\
\til b &  \til d
\end{array} \right), \nonumber\\ \label{psdgroup}
\end{eqnarray}
composed as a semi-direct product of one specific ``dilatation'', of ${\mathrm{Det}} \, G_{dil}=1/d$, and the special conformal transformation\footnote{remember that one can always realize the special conformal transformation as a product of tree consecutive transformations --- inversion, translation by $b/d$ and one more inversion.} of parameter $b/d$ --- with the well known group laws: $d_3=d_1d_2$ and $b_3=b_2d_1+b_1/d^2_2$. Notice that, differently from the SD transformation (i.e. the simple inversion), the inverse element $G^{-1}_{psd}$ in the PSD case \emph{is not coinciding} with $G_{psd}$. It is instead  providing  a  group-theoretical meaning of the duality transformations (\ref{dual param. gmn}) for the parameters of the superpotential, that according to our general duality formula (\ref{int s til}) are parametrizing the group of the duality transformations: $d \til\s/d \s = - \ka L_{gr} \, \til W(\til \s)$.  Hence the parametric form of the partially self-dual superpotential (\ref{W dmn}) is determined by the PSD duality group elements $G_{psd}(b,d)\in GL(2,R)$. 
Thus, our particular choice of the PSD superpotential (\ref{W dmn}) introduces certain group structure on the space of W-parameters, representing  the set of couplings in the potential $V(\s,G_{pds})$ of the 3d matter field of the NMG-matter model. The superpotential $\til W(\til \s, G^{-1}_{psd})$ of the  second member of the dual pair of NMG models is then parametrized by the corresponding inverse elements $G^{-1}(b,d)$. This is in fact the NMG \emph{gravitational counterpart} of the $d=2$ QFT's  self-duality requirements. It is also in the origin of the important property of the partially self-dual models, namely  that the $\beta-$ functions  of such pairs of models have the same form, i.e. the PSD transformations (\ref{til sigma gmn}) and (\ref{psdconf}) are keeping invariant the form of the corresponding RG equation:
\begin{eqnarray}
\frac{d\s}{dl} = \frac{6\epsilon B (\s-\s_a)}{\kappa^2 (B(\s-\s_a)^2+D)}\big(1-\ka^2 L^2_{gr}(B(\s-\s_a)^2+D)^3\big)=-\beta_{psd}(\s;\s_a,B,D),    \label{rgsd}
\end{eqnarray}
but with its W-parameters $\s_a, B,D$  replaced by their duals: $\til \s_a$, $\til B$ and $\til D$. Thus, the RG's equation of the dual $pCFT_2$ has the expected form $d\til \s/dl=-\beta_{psd}(\til \s;\til \s_a,\til B,\til D)$. The ''slight'' difference between the \emph{invariance conditions} of the RG equations and of forms of the $\beta$-functions of the considered SD and PSD models has its origin in the different group properties  of their coupling transformations (\ref{sdu}) and (\ref{til sigma gmn}).

The specific ``fractional-linear'' form of the PSD transformation (\ref{psdconf}) requires further investigation of the problem  of whether  strong couplings $u$ are mapped to weak ones $\til u$  for all the values of the parameters $b$ and $d$. Let us first note one particular feature of our PSD transformation (\ref{psdconf}), namely  that $ u(0)=0=\til u(0)$ and $u(\infty)=1/bd^2=\til u(1/bd^2)$ and therefore it is mapping the the positive semi-axis  $u\in (0,\infty)=R_+$ to the finite interval $\til u \in (0,1/bd^2)$. It is then clear that  in order to transform the large values of $u$ ( and of $\s$ as well) into the small ones of the $\til u$ and vice-versa we have to impose the following restriction on the values of  the parameters $b$ and $d$:
\be 
 bd^2\gg 1 \quad\quad \text{or equivalently} \quad\quad BD^2\gg \frac{1}{\ka^2 L^2_{gr}}=2|m^2|/\ka^2 . \label{strwe}
\ee

\noindent
\textit{Symmetries of RG equations vs. Duality.} As we have shown, the ``duality invariance'' of the RG equations and of the form of the holographic $\beta$-functions turns out to be one of the main  features  specific for the class of the SD and PSD models only\footnote{ in the case of generic duality transformations (\ref{def dual}) and (\ref{dutrans}), the pairs of dual $\beta-$functions are related by the eq.(\ref{dual beta}) and the corresponding RG equations does not remain invariant.}. It is important however to  mention that the SD and PSD duality transformations are \emph{not exhausting} all the symmetries of the RG equation. In fact one can find more symmetries of the corresponding RG equations, that  are \emph{not preserving} neither the central function nor the free energy. Therefore the invariance of RG equations under a kind of strong-weak coupling transformations   \emph{can't be considered} as a definition of (partial) self-duality of $pCFT_2$'s under investigation. We shall give   an example of such ``additional'' symmetries of the RG eq.(\ref{rgsd})for the SD model. Let us first rewrite it in the following equivalent form :
 \begin{eqnarray}
\frac{dg}{dl}=g^2-a^2, \quad\quad g=8B^2L^2_{gr}\s^2,\quad\quad a=\frac{8BL_{gr}}{\ka}.\label{eq.g}
\end{eqnarray}
Apart of the already discussed duality symmetry $\til g =\frac{a^2}{g}$, it is also invariant under specific fractional linear transformations
\begin{eqnarray}
g(l)\rightarrow g'(l)=\frac{\cosh(a\gamma)g(l)-a \sinh(a \gamma)}{-\frac{\sinh(a\gamma)}{a}g(l)+\cosh(a\gamma)},\label{uly}
\end{eqnarray}
where $\gamma\in R$ is an arbitrary real parameter. These transformations\footnote{notice that the corresponding transformations of  the original ``coupling variable'' $\s=\frac{\sqrt{g}{2}}{2a\ka}$ are also forming an $SO(1,1)$ group} can be recognized as an $SO(1,1)$ subgroup of $SO(2,1)$. In spite of the fact that for certain restrictions on the parameters $a$ and $\gamma$ they are mapping strong to weak couplings, they \emph{are not} keeping invariant the corresponding central 
functions,anomalous dimensions and free energy and therefore are not representing duality transformations at all.

It is worthwhile to  mention that the above eq.(\ref{eq.g}) also appears as RG equation for two rather different models:(a)the RG eq. of the $pCFT_2$ dual to the NMG model of linear superpotential (see ref.\cite{nmg}); (b) the well known one-loop RG equation  with the perturbative $\beta$-function given by (\ref{pertrg}), specific for the perturbations of the so called $\Phi_{13}$ relevant operators of the minimal $CFT_2$'s \cite{cardy,x}. In both cases however  neither its inversion symmetry $\til g =\frac{a^2}{g}$ nor the above considered $SO(1,1)$ symmetry (\ref{uly}) act as proper strong-weak coupling duality.

\textit{Few comments and relevant open questions:}

$\bullet$ The two simple examples of self-dual and partially self-dual superpotentials, that generate very specific (limits of) duality groups and give rise to  self-dual pCFT$_2$'s, are indeed not representing all the possible  (partial) self-duality transformations. One could considerer, for example, a simple three  parameters  quadratic superpotential, that turns out to generate (within the NMG context considered in this section) more general $SL(2,R)$ duality transformations.

$\bullet$ The most interesting cases of explicit realizations of the self-duality in the mentioned 2d and 4d QFT's models(see for example \cite{seiberg,seibit,cardy-itz}), that have the $SL(2,Z)$ (sub)group as duality symmetries, are known to be with complex valued coupling constant (or equivalently of two real couplings). In the case of the considered NMG-matter models,  it will corresponds to specific two scalar fields matter interactions superpotentials. The problem of the generalizations of the concepts of NMG duality (\ref{def dual}) to the case of complex fields, based on an appropriate I-st order system of DW's equations, and of the corresponding constructions of the two $\beta-$functions  in terms of these superpotential is under investigation.

\setcounter{equation}{0}
\section{Holographic RG flows and self-duality} \label{Holographic RG flows}

The off-critical NMG$_3$/QFT$_2$ conjecture, based on the holographic RG eqs.(\ref{rg}), is a natural generalization of the standard ($m^2\rightarrow \infty$) holographic RG \cite{VVB,rg}. Let us remind its content: there exists a family of QFT$_2$ such that their near-critical behaviour and phase structure admit a non-perturbative geometrical description in terms of DW's solutions of the NMG-matter model (\ref{acaoo}) with an appropriately chosen superpotential $W(\sigma)$. The first part of this statement concerns the identification of the  NMG vacua $(\sigma^*_A,L_A,y_A)$ with the critical $CFT_2$-data of the dual QFT$_2$ as we have done in Sect. 2 above. Its second part is about the explicit relation between the set of ``consecutive"  DW solutions 
$$DW_{k,k+1}=\Big(\sigma(z),e^{\varphi(z)};z\in R \quad|\quad\sigma^*_k,L_k\rightarrow \sigma^*_{k+1},L_{k+1}\Big),\;\;\;\;\sigma \in R,$$ 
and all the $QFT_2$ phases $p^{ml}_{k,k+1}=(\sigma^*_{k}(IR),\sigma^*_{k+1}(UV))$ described by the coupling constant $\sigma_{k,k+1}(l)$ and the s.p. of the free energy $F_s(\sigma)\approx e^{-\varphi(\sigma)}$ behaviours. In what follows, our  attention is 
concentrated on the properties of the couples of neighbour DW's of common boundary ($\sigma^*_{\uv},L_{\uv},y_{\uv}$) that have different (IR)-horizons b.c.'s, say for example $(\sigma^*_{\ir},\sigma^*_{\uv})$ and $(\sigma^*_{\uv},\infty)$. They represent the main ingredient in the description of the phase transitions and of the nature of the holographic RG flows \cite{nmg,oldholo}.


\subsection{The phases of the self-dual superpotential}

 Let us  recall which of the solutions of the RG  eqs. (\ref{rg}) and (\ref{fs}) -- defined within a given interval, say $\sigma\in (\sigma_{+},\infty)$ or $\sigma\in (\sigma_{-},\sigma_s)$, etc. -- can be identified as describing the particular \emph{massive RG flows} in the related QFT$_2$. The main requirement is that the running coupling $|\sigma(l)-\sigma_{+}|$ gets its maximal value for a \emph{finite} RG distance, for example $\sigma(L_{max})=\infty$ or $\sigma(L_{max})=\sigma_{max}=|\sigma_s - \sigma_+ |$, etc., imposing that the correlation length, say $\xi(\infty)=\xi_{max}= 1/M_s$ always has a finite maximal value. Then its inverse defines the smallest  mass gap in the energy spectrum, and as a consequence of eqs. (\ref{fs}) the corresponding 2-point correlation function manifests an \emph{exponential decay}: $e^{-M_{ms}|x_{12}|}$ -- typical for the IR limit of the propagator of a free massive particle. This behaviour has to be compared to the one corresponding to the \emph{massless RG flows}, where the maximal distance $|\sigma_{\ir}-\sigma_{\uv}|$ from the starting (at $L_*=0$) UV critical point is reached for $L_{max}=\infty$, i.e. $\xi(L_{max})= \infty$ and therefore no mass gap exists, since $M^2=0$. As a result, the correlation functions at an IR critical point have power-like (scale invariant) behaviour.  
 
 Examples of such massless phases are found in the self-dual superpotential $W=B\sigma^2$. Taking $B>0$, we have two massive phases $p_{{\mathrm{flat}}}^\ms=(0,\sigma_{+})$ and $p_{{\mathrm{n.s.}}}^\ms =(\sigma_{+},\infty)$, described holographically by two DW's, one of $E_3/AdS_3$ type and the other of AdS$_3$/n.s. type, with a common boundary at the type (b) vacuum $\sigma_+ = 1 / \sqrt{\ka L_{gr}B}$. We consider here only positive values of $\s$ because of the $Z_2$ symmetry of the superpotential. This massive nature of the phases can be apprehended by the correlation length $\xi(\s)$, which can be found through the corresponding  RG equation: 
\begin{eqnarray}
\frac{d\sigma}{dl}=-\beta_{qp}(\sigma)= \frac{4\epsilon}{\kappa^2\sigma}\big(1-\kappa^2L^2_{gr}B^2\sigma^4\big) .    \label{qpot}
\end{eqnarray}
It has as solution $\sigma^2(l)=\sigma^2_{+} \, \coth(l_0-\frac{y_{+}l}{2})$, leading to
\begin{eqnarray}
e^{-l}\approx\xi(\sigma) = \left[ \frac{(\sigma^{2}/\s_+^2) + 1}{(\sigma^{2}/\s_+^2) - 1} \right]^{\frac{1}{y_{+}}} \left[ \frac{(\sigma^{2}_0 / \s_+^2) - 1}{(\sigma^{2}_0 / \s_+^2) + 1} \right]^{\frac{1}{y_{+}}},\quad y_{+}=-\frac{16\epsilon BL_{gr}}{\kappa}. \label{flatsol}
\end{eqnarray}
This expression is singular at $\s_+$, and the divergence (for $\epsilon = -1$) of the scale factor shows that it is an UV vacuum. On the other hand, the correlation length takes \emph{finite} values at both the singular point $\sigma_s = 0$, which is a flat vacuum in the weak-coupling ``massive-flat" phase, and at $\sigma \to \infty$, in the standard strong-coupling massive phase. It is easy to calculate the corresponding mass gaps, say
$$ M_{{\mathrm{n.s.}}}(\sigma_0)=1/\xi(\infty)= \big(\kappa L_{gr}B\sigma^{2}_0-1\big)^{\frac{1}{y_{+}}}\big(\kappa L_{gr}B\sigma^{2}_0+1\big)^{-\frac{1}{y_{+}}} , $$
thus confirming the massive nature of $p_{{\mathrm{flat}}}^\ms = (0,\sigma_{+})$ and $p_{{\mathrm{n.s.}}}^\ms = (\sigma_{+},\infty )$. 
The duality transformation here is known from Sect.\ref{self-duality} to be $\til \s = \s_+^2/ \s$,  leaving the superpotential invariant: $\til W(\til\s) = B \til\s^2$, as well as the vacuum $\s_+ = \til\s_+$. But the singular points are ``exchanged'' through $\til\s (\s_s = 0) = \infty$ and $\til\s_s (\s \to \infty) = 0$, and so there is a correspondence between the two massive phases with strong and weak coupling:  $M_{{\mathrm{n.s.}}}(\sigma_0)=M_{{\mathrm{flat}}}(\til\sigma_0)$.

\subsection{Phase transitions and partial self-duality}

The phase structure of the partially self-dual superpotential  $W(\s) = [B (\s -\s_a)^2 + D]^{3/2}$, studied in Sect.3.2., depend on the range of the values of the  parameters $B$ and $L_a^{-1} = \ka D^{3/2}$, as shown on  fig.1. In the case (I.a), corresponding to $L_a > L_{gr}$ and 
$B> 0$, we have an IR critical point at $\s_a$ and two UV critical points at $\s_b^\pm$. There are four DW's, which  describe the four different phases of the corresponding dual $QFT_2$ in this region of the parameter space:
\br
p^\ms_{\ns} = (-\infty \; , \; \s_b^-) \; ; \;\;\;\; p^\ml_{\uv/\ir} = (\s_b^- \; ,\; \s_a) \; ; \;\;\;\; p^\ml_{\ir/\uv} = (\s_a \; , \; \s_b^+) \; ; \;\;\;\; p^\ms_\ns = (\s_b^+ \; , \; \infty) .
\er
The nature -- massive (ms) or massless (ml) -- of the phases can be easily read from the scale factor's (\ref{scale factor dmn}) analytic properties, which determines the correlation length of the dual $pCFT_2$:
\be
\xi(\s) \approx  \left( \frac{\s_0 - \s_a}{\s - \s_a} \right)^{\frac{1}{y_a}}  \left( \frac{\s_0 - \s_b^+}{\s - \s_b^+} \right)^{\frac{1}{y_+}} \left( \frac{\s_0 - \s_b^-}{\s - \s_b^-} \right)^{\frac{1}{y_-}} \prod_{i=2}^3 \left[ \frac{(\s_0 - \s_a)^2 + (D - g_i) / B}{(\s - \s_a)^2 + (D - g_i) / B} \right]^{- x_i} . \label{correl length dmn}
\ee
The critical exponents are given by eqs.(\ref{A23}). They also satisfy the remarkable  NMG ''resonance'' condition 
$$\frac{1}{y_a}+\frac{1}{y_+}+\frac{1}{y_-}=\sum_{i=2}^{3}x_i,$$ 
that turns out to hold  for all the QFT's models, obtained by  $NMG_3$ holography \cite{nmg}. The ``initial condition'' $\s_0 \equiv \s|_{l = 0}$ of RG rescaling  can be further fixed by requiring that $L_*^{(0)} \approx 1$.
As we have shown in Sect.\ref{Examples of dual and self-dual}, for $\epsilon = -1$ we have $y_a < 0$ and consequently $\xi(\s_a) \to 0$; therefore $\s_a$ is an IR critical point, while for the (b) type critical points : $y_\pm > 0$ hence $\xi(\s_B^\pm) \to \infty$ and $\s_b^\pm$ are UV critical points. Notice that the finite values of $\xi(\s)$ when $\s \to \pm \infty$  and, as a consequence, the existence and  properties of the massive phase are due to the above mentioned NMG resonance condition, i.e. the fact that the sum of the critical exponents $\nu_k$ (of all the critical points) vanishes. The corresponding values of the mass gaps for the massive phases can be evaluated at these limits $\s \to \pm \infty$, which correspond to naked singularities in the NMG-geometry. For example,  the strong-coupling massive phase $p^\ms_\ns = (\s_b^+ \; , \; \infty)$, is characterized by the asymptotic value of the correlation length (\ref{correl length dmn}), which determines the smallest mass in the dual model:  
\be 
M_{(\ms)} \approx \xi^{-1}|_{\s \to \infty} =  \left( \s_0 - \s_a \right)^{\frac{1}{y_a}}  \left( \s_0 - \s_b^+\right)^{\frac{1}{y_+}} \left( \s_0 - \s_b^- \right)^{\frac{1}{y_-}} \prod_{i=2}^3 \left[ (\s_0 - \s_a)^2 +(D - g_i)/B \right]^{-x_i}
\ee

We next describe the duality between the strong- and weak-coupling  phases of the considered partially self-dual pCFT$_2$ model, i.e. how the duality transformation (\ref{til sigma gmn})-(\ref{dual param. gmn}) is effectively mapping the phases of this model. As we have demonstrated in Sect.3.2., the phases \emph{duals} of the above considered (I.a) case are those of the (II.b)- model (see fig.2.):
\br
p^\ms_{flat} = (\til\s_M^-  ,  \til\s_b^-) \; ; \;\;\;\; p^\ml_{\uv/\ir} = (\til\s_b^- ,\til \s_a) \; ; \;\;\;\; p^\ml_{\ir/\uv} = (\til\s_a , \til \s_b^+) \; ; \;\;\;\; p^\ms_{flat} = (\til\s_b^+  , \til \s_M^+) .
\er
i.e. of our original partially self-dual model, but now with different range of the values of the parameters: $\til B < 0$ and $\til L_a < L_{gr}$. The correlation length $\til\xi(\til\s)$ has the same form (\ref{correl length dmn}) as above, but with the parameters exchanged by the duality according to eq.(\ref{dual param. gmn}). Notice that although $B$ changes its sign, the critical exponents do not, since the ratio $L_{gr}/\til L_a$ is now greater than unity.  We have to remind that the corresponding ''dual massive'' phases correspond to non-singular, E$_3$/AdS$_3$ DW's solutions, with a mass gap given by 
\br
& \til M_\ms \approx \til \xi |_{\til \s \to \til \s_M^+} = \left( \frac{\til\s_M^+ - \til\s_a}{\til\s_0 - \til\s_a} \right)^{- \frac{1}{\til y_a}}  \left( \frac{\til\s_M^+ - \til\s_b^+}{\til\s_0 - \til\s_b^+} \right)^{-\frac{1}{\til y_+}} \times \nonumber \\
& \quad\quad\quad\quad\quad\quad \times
 \left( \frac{\til\s_M^+ - \til\s_b^-}{\til\s_0 - \til\s_b^-} \right)^{-\frac{1}{\til y_-}} \prod_{i=2}^3 \left[ \frac{(\til\s_M^+ - \til\s_a)^2 + (\til D -  g_i) / \til B}{(\til \s_0 - \til\s_a)^2 + (\til D - g_i) / \til B} \right]^{\til x_i} . \nonumber
\er
while in the (Ia) case they are related to the singular $AdS_3/n.s.$ DW's, interpolating between one $AdS_3$ vacua and a naked singularity. The large values of the formerly unbounded coupling $\s$  is now mapped at the (small) finite values of $\til \s$ in the neighbours of the Minkowski vacua\footnote{ Notice that such vacua of $W(\s_M)=0$ are \emph{not} representing
 ''conformal critical'' points, but instead are defining a particular massive phase \cite{nmg},\cite {oldholo}.}. We can conclude that in the dual theory the ``infinitely strong'' couplings are mapped into a finite values, both however  corresponding to massive phases: hence  the strong coupling massive phase is mapped to certain ''dual'' weak coupling massive phase. The dual massless phases, on the other hand, are ``stretched'' by the duality transformation, as one can see from eq.(\ref{til s b dmn}): the interval $(\til\s_a, \til\s_b^\pm)$ is ``longer'' than its dual, for $L_a/L_{gr} > 1$.

Similar statements are valid for all the other pairs of dual models described in Sect.3.2.:
$$ {\mathrm{(I.a)}} \Leftrightarrow {\mathrm{(II.b)}} \; ; \;\;\;  {\mathrm{(I.b)}} \Leftrightarrow {\mathrm{(II.a)}} \; ;  \;\;\;  {\mathrm{(III.a)}} \Leftrightarrow {\mathrm{(III.b)}}$$
Let us mention that the behaviour of the correlation length and the properties of the marginally degenerate cases $(III.a)$ and $(III.b)$, that in fact describe a pair of dual models with an infinite order phase transition at the critical point $\s_{a} = \s_b$ and having two massive phases, are quite similar to the ones of the NMG model of quadratic superpotential, studied in ref.\cite {oldholo}.

Few comments are now in order:

(a) the phase structure, the corresponding RG flows and the duality  relations between different phases of our second example of partially self-dual ''periodic'' superpotential (\ref{W cos})(we have introduced in App.A.), are rather similar to the one we have described in this subsection;

(b) the holographic RG flows in the pCFT$_2$ model dual to the NMG of quadratic superpotential
\be
W(\s) =  B (\s -\s_a)^2 + D ,\;\;\;\;\;\;\; D\neq 0
\ee
can be easily found by applying the methods developed in Sect.3.1. and by using  the results of refs. \cite{nmg}, \cite{oldholo}. Although (for $D\neq 0$) it is neither self-dual (as in the $D=0$ case) nor partially self-dual, it possess a rich and interesting phase structure \cite {nmg,oldholo}. It is worthwhile to also mention the well known fact that it represents the near-critical behaviour of an arbitrary (even) superpotential.


\setcounter{equation}{0}
\section{Discussion}

The holographic RG methods, when applied to the NMG-matter models with appropriate superpotentials, provide important critical (about certain CFT$_2$'s) and off-critical (of the corresponding pCFT$_2$'s) data, which can be used for their identification with the --- already known --- perturbative and exact CFT$_2$ and pCFT$_2$'s results \cite{x, bpz, fat,gms}. It is worthwhile to  remind once more that all the information about the holographic RG flows and phase transitions in the QFT$_{2}$'s dual to the NMG model (\ref{acaoo}) are not sufficient for the complete identification of the pCFT$_2$ dual to a given  NMG-matter model. One has to further consider the difficult problem of the construction of the off-critical correlation functions of 2d fields dual to the 3d matter scalar, by studying the linear fluctuations of the metrics and of the scalar field around the DW's solutions \cite{gub, rg, japa, 8}. The real problem with the verification of the validity of the off-critical $(a)AdS_3/pCFT_2$  conjecture consists however in the comparison of the \emph{strong-coupling holographic} results, based on the exact $\beta$-functions, with the known \emph{perturbative}, near-critical calculations of the corresponding 2d models \cite{azz, cardy, x, fat, gms}. The construction of a particular class  of strong-weak coupling self-dual pCFT$_2$'s models, i.e. the holographic duals of selected pairs of NMG-matter models with partially self-dual superpotentials, described in Sects.\ref{Duality} and \ref{Holographic RG flows},  represents an important exception. In this case it becomes possible to compare the  holographic non-perturbative results with the ones obtained by the conformal perturbation theory \cite{x,cardy}.

Another important problem concerning the $(a)$AdS$_3$/pCFT$_2$ correspondence, in the particular case of the NMG model (\ref{acaoo}), is related to the \emph{negative values} of the central charges (\ref{ch}) for $\epsilon=-1$ and $m^2<0$. These are  usually interpreted as \emph{non-unitary} CFT$_2$'s. Let us assume that all these CFT$_2$'s, without any extra symmetries present, are described by the representations of two  commuting Virasoro algebras, characterized by their central charges $c_L = c_R = c$, and the set of scaling dimensions and spins \cite{bpz}. In all the cases when $ c<0$, the corresponding CFT$_2$'s contain primary fields (states) of negative dimensions (and negative norms), and hence they represent non-unitary QFT$_2$'s \footnote{Some of them turn out to describe interesting 2d statistical models, as for example the one of central charge $c=- 22 / 5$, known as  Lee-Yang edge singularity \cite{muss}.}. As it is well known, in the interval $0<c<1$  there exists  an infinite series of ``minimal" \emph{unitary quantum models} corresponding to
$c^-_{{\mathrm{quant}}}(p)=  1 - 6 Q_p^2$, with $Q_p= \sqrt{\frac{p+1}{p}} -\sqrt{\frac{p}{p+1}}$ and $ p=3,4,5,...$, while 
the models with $c > 25$ give rise to unitary representations used in the quantization of the Liouville model \cite{azz}: $c_{+}(b) =  1 + 6(b+\frac{1}{b})^2$, where the parameter $b$ is related to the Liouville coupling constant. On the other hand, the derivation of the Brown-Henneaux \cite{9} central charge formula $c=\frac{3L}{2G}$, as well as its NMG generalizations (\ref{ch}), are based on the ``Dirac quantization'' of the classical Poisson brackets of the  Virasoro algebra, and by further identifying the classical central charge $c_{{\mathrm{class}}}$ for $L\gg l_{pl}$ with the ``quantum'' central charge $c_{{\mathrm{quant}}}$ of the ``dual'' boundary CFT$_2$. The well known fact, coming from the standard procedure of the Liouville models \cite{azz} and  of the ``minimal'' models quantizations \cite{fat}, is that this classical central charge is receiving quantum corrections, i.e. starting  from the $c_{{\mathrm{class}}}^{\pm} = \pm 6b^2$ we are getting their  ``corrected'', exact values $c_{{\mathrm{quant}}}^{\pm}=1\pm 6(b\pm\frac{1}{b})^2$. 
In the classical limit $\hbar\rightarrow 0$ one obtains $c_{{\mathrm{quant}}}^-\rightarrow c_{\mathrm{class}}^-\approx-\infty$, i.e. the corresponding classical (and semiclassical) central charges are very big, \emph{negative} numbers \cite{fat}. Similarly, for the  limits of the central charges of the Liouville's model \cite{azz},  we have $c_{\mathrm{class}}^+\approx\infty$. Hence the classical (and semi-classical) large negative central charges are a common feature  of  all the $c^-_{\mathrm{quant}} <1$ models and of their supersymmetric $N=1$ extensions. It is therefore  important to bear in mind that given the values of the (semi-)classical limits of the central charges of certain class of CFT$_2$'s, further investigations of the limiting properties of the anomalous dimensions of the primary fields are also needed, in order to conclude whether such 2d CFT's  belong to the non-unitary ($c^-_{\mathrm{quant}}<0 $) case, or else to the interval $0<c^-_{\mathrm{quant}}<1$,  where unitary models are known to exist.

Our final comment concerns  the eventual higher dimensional $d>3$ generalizations  of the duality concepts and of the specific examples we have considered in the present paper. It should be stressed that the presence of the $R^2$ terms (specific for the NMG gravity) and the knowledge of the corresponding I-st order system of eqs.(\ref{sis}) were \emph{essential} in the derivation of our 3d NMG duality conditions (\ref{def dual}). Due to the specific form of the NMG central function, it is clear that the pure EH action coupled to scalar matter, and the corresponding dual pCFT$_{d-1}$, do not provide  examples of dual and self-dual models (even in the 3-dimensional case); at least not in the context proposed in Sect.\ref{Duality} above. Therefore one has to look for appropriate higher dimensional  ``higher curvature'' gravitational actions of Lovelock type, as for example the ones containing the Gauss-Bonnet term and/or specific combinations of cubic or quartic powers of the curvature tensors similar to the actions of  Quasi-Topological gravities \cite{myers, oliva, mann}. As in the case of 3d NMG models studied in the present paper, the main ingredients of such holographic duality constructions are again the explicit forms of the corresponding  $a$- and $c$-central functions, of the exact $\beta$-functions and of the holographic free energy. There exist many indications of how one can formulate an appropriate generalization of the considered  NMG duality conditions in certain higher dimensional models, for which the holographic RG methods, based on the DW's solutions and on the first order order system of equations \cite{lovedw, loverg} are well established.  Our preliminary results \cite{sdual} provide convincing arguments that the NMG-like duality conditions (\ref{def dual}) can be realised only in a very particular class of  higher dimensional gravity models: For  $d=4$, i.e. for the construction of self-dual pCFT$_3$'s, the appropriate model allowing such partial self-dualities is  the $d=4$ \emph{cubic} Quasi-Topological gravity \cite{oliva, lovedw, loverg}; while for the $d=5$ case it turns out to be  the recently constructed \emph{quartic} Quasi-Topological Gravity, with the linear and the quartic terms only \cite{mann}.


\section{Appendix. Partially self-dual NMG's with periodic superpotential}
\label{App}

The vacua structure of the following superpotential 
\be
W(\s) = B \left[ D - \cos (\a \s) \right] , B < 0  \label{W cos} 
\ee
consists in two type (a) vacua at $\s_a^{(0)} = 0$ and $\s_a^{(\a)} = \pi /\a$ and few type (b) ones (within the interval $\s\in (0, \pi/\a)$). We can further rewrite the parameters $B$ and $D$ in an equivalent form as
\be
B = - \frac{L_0 - L_\a}{2 \ka L_0 L_\a} \; , \;\; D = \frac{L_0 + L_\a}{L_0 - L_\a} ,
\ee
by introducing an obvious notation for the vacua scales $L_{0, \a}$.
The condition $B < 0$, i.e. $L_0 > L_\a$, implies that $D > 1$, hence Minkowski vacua or Janus-type geometries are excluded.

Using (\ref{int s til}), we have
\be
\tan \left[ \frac{B \a \ka L_{gr} \sqrt{ D^2 - 1} \, \til\s}{2} \right] = \sqrt{\frac{D+1}{D-1}} \, \tan \left[ \frac{\a \s}{2} \right] . \label{s til cos}
\ee
This gives:
\be
\til W(\til\s) = \til B \left[ \til D - \cos (\til \a \til\s) \right] ,
\ee
where
\be
\til B = - \frac{1}{\ka^2 L_{gr}^2 B ( D^2 - 1)} \; , \;\; \til D = - D \; , \;\; \til\a = B \ka L_{gr} \sqrt{D^2 - 1} \, \a . \label{parameters cos dual}
\ee
Thus, we see that the case considered: $B < 0$, $D > 1$, is dual to other case: $\til B > 0$, $\til D < -1$. We can integrate the scale factor, to find
\be
e^{\p (\s)} = e^{\p_0} (1 + \cos \a\s)^{x_1} \, (1 - \cos \a\s)^{x_2} \, |\delta_+ - \cos \a \s |^{x_3} \, |\delta_- - \cos \a \s |^{x_4} , \label{scale factor cos}
\ee
where
\br
&& x_1 = - \frac{L_0 L_\a^2}{2 \a^2 \left[ L_0 - L_\a \right] \left[  L_\a^2 - L_{gr}^2 \right] }  , \;\;   x_2 = \frac{L_\a L_0^2}{2 \a^2 \left[ L_0 - L_\a \right] \left[  L_0^2 - L_{gr}^2 \right] } , \\ \label{x1 cos}
&& x_3 =  \frac{L_0  L_\a }{4 \a^2 \left[ L_{gr} + L_\a \right] \left[ L_0 + L_{gr} \right] } , \;\;  x_4 = \frac{L_0  L_\a }{4 \a^2 \left[ L_0 - L_{gr} \right] \left[ L_\a - L_{gr} \right] } , \\
&& \delta_\pm = \frac{1}{\left( L_0 - L_\a \right)}  \left[ L_0 + L_\a \pm 2 \frac{L_0 L_\a}{L_{gr}} \right] . \label{delta cos}
\er
The condition for the existence of a DW solution connecting two type (a) vacua, i.e. the condition for the absence of singularities of the scale factor for $\s \in (\s_a^{(0)} , \s_a^{(\a)} )$, is that $\delta_+ > 1$ and $\delta_- < -1$, implying $L_{0,\a} > L_{gr}$, thus
$$0 < L_{gr} < L_\a < L_0 . $$ 
In this case, we have a DW connecting a boundary at $\s = 0$ and a horizon at $\s = \pi/\a$.

The description of its phase structure, the nature of the phase transitions  as well as the duality relations between the different phases (for different ''dual'' values of the superpotential parameters), following the methods developed in Sect.3.2.2. and Sect.4.2, is straightforward.


\end{document}